\pdfoutput=1

\documentclass[11pt,twoside,a4paper,cmspaper,final,collab]{cms-tdr}
\def\svnVersion{77706b4-D}\def\svnDate{2019/07/28}\def\cmsCernNoTag{CERN-EP-2019-025}\def\cmsCernDate{\today}\def\cmsMessage{"Published in the Journal of High Energy Physics as \href{http://dx.doi.org/10.1007/JHEP10(2019)142}{\doi{10.1007/JHEP10(2019)142}.}"}

\begin{document}\cmsNoteHeader{HIG-18-014}

\hyphenation{had-ron-i-za-tion}
\hyphenation{cal-or-i-me-ter}
\hyphenation{de-vices}
\RCS$HeadURL$
\RCS$Id$

\newlength\cmsFigWidth
\ifthenelse{\boolean{cms@external}}{\setlength\cmsFigWidth{0.49\textwidth}}{\setlength\cmsFigWidth{0.65\textwidth}}
\ifthenelse{\boolean{cms@external}}{\providecommand{\cmsLeft}{upper\xspace}}{\providecommand{\cmsLeft}{left\xspace}}
\ifthenelse{\boolean{cms@external}}{\providecommand{\cmsRight}{lower\xspace}}{\providecommand{\cmsRight}{right\xspace}}
\newcommand{\pb}{\ensuremath{\unit{pb}}\xspace}
\newcommand{\fb}{\ensuremath{\unit{fb}}\xspace}
\newcommand{\ptldgtrk}{\ensuremath{\pt^{\text{track}\xspace}}}
\newcommand{\ptmisscalo}{\ensuremath{\pt^\text{miss,calo}}\xspace}
\newcommand{\tanbeta}{\ensuremath{\tanb}\xspace}
\newcommand{\mHpm}{\ensuremath{m_{\PH^\pm}}\xspace}
\newcommand{\mHpmLight}{\ensuremath{\mHpm<\mtop-m_{\cPqb}}\xspace}
\newcommand{\mHpmIntermediate}{\ensuremath{\mHpm\sim\mtop}\xspace}
\newcommand{\mHpmHeavy}{\ensuremath{\mHpm>\mtop-m_{\cPqb}}\xspace}
\newcommand{\mtop}{\ensuremath{m_{\cPqt}}\xspace}
\newcommand{\Rtau}{\ensuremath{R_\Pgt}\xspace}
\newcommand{\Rtaudef}{\ensuremath{\ptldgtrk/\pt^{\tauh}}\xspace}
\newcommand{\RtauLess}{\ensuremath{R_\Pgt<0.75}\xspace}
\newcommand{\RtauMore}{\ensuremath{R_\Pgt>0.75}\xspace}
\newcommand{\Rbbmin}{\ensuremath{R_{\text{bb}}^\text{min}}\xspace}
\newcommand{\Hplus}{\ensuremath{\PH^+}\xspace}
\newcommand{\Hminus}{\ensuremath{\PH^-}\xspace}
\newcommand{\taunu}{\ensuremath{\Pgt^{\pm}\Pgngt}\xspace}
\newcommand{\Hpmtaunu}{\ensuremath{\PH^{\pm}\to\Pgt^{\pm}\Pgngt}\xspace}
\newcommand{\heavyProduction}{\ensuremath{\Pp\Pp\to\cPqt\cPqb\PH^{\pm}}\xspace}
\newcommand{\BtopToHpm}{\ensuremath{\mathcal{B}(\cPqt\to\cPqb\PH^{\pm})}\xspace}
\newcommand{\DY}{\ensuremath{\cPZ/\gamma^{*}}\xspace}
\newcommand{\jetToTauh}{\ensuremath{\text{jet}\to\tauh}\xspace}
\newcommand{\tauhjets}{\ensuremath{\tauh}~+~jets\xspace}
\newcommand{\ltauh}{\ensuremath{\ell}~+~\ensuremath{\tauh}\xspace}
\newcommand{\lnotauh}{\ensuremath{\ell}~+~no~\ensuremath{\tauh}\xspace}
\newcommand{\sigmaHplus}{\ensuremath{\sigma_{\PH^\pm}}\xspace}
\newcommand{\BHpmtaunu}{\ensuremath{\mathcal{B}(\Hpmtaunu)}\xspace}
\newcommand{\heavyLimit}{\ensuremath{\sigmaHplus\BHpmtaunu}\xspace}
\newcommand{\mhmodm}{\ensuremath{m_{\Ph}^\text{mod-}}\xspace}
\newlength\cmsTabSkip\setlength\cmsTabSkip{3pt}
\ifthenelse{\boolean{cms@external}}{\providecommand{\CL}{C.L.\xspace}}{\providecommand{\CL}{CL\xspace}}
\providecommand{\cmsTable}[1]{\resizebox{\textwidth}{!}{#1}}

\cmsNoteHeader{HIG-18-014}
\title{Search for charged Higgs bosons in the $\PH^{\pm} \to \tau^{\pm}\nu_\tau$ decay channel in proton-proton collisions at $\sqrt{s}=13\TeV$}

\date{\today}

\abstract{A search is presented for charged Higgs bosons in the $\PH^{\pm} \to \tau^{\pm}\nu_\tau$ decay mode in the hadronic final state and in final states with an electron or a muon. The search is based on proton-proton collision data recorded by the CMS experiment in 2016 at a center-of-mass energy of 13\TeV, corresponding to an integrated luminosity of 35.9\fbinv. The results agree with the background expectation from the standard model. Upper limits at $95\%$ confidence level are set on the production cross section times branching fraction to $\tau^{\pm}\nu_\tau$ for an H$^{\pm}$ in the mass range of 80\GeV to 3\TeV, including the region near the top quark mass. The observed limit ranges from 6\pb at 80\GeV to 5\fb at 3\TeV. The limits are interpreted in the context of the minimal supersymmetric standard model \mhmodm scenario.}

\hypersetup{
pdfauthor={CMS Collaboration},
pdftitle={Search for charged Higgs bosons in the H+- -> tau+-nu decay channel in proton-proton collisions at sqrt(s)=13 TeV},
pdfsubject={CMS},
pdfkeywords={Higgs bosons, charged Higgs bosons, Hplus, Higgs sector extensions, 2HDM, MSSM, taus, muons, electrons}}

\maketitle

\section{Introduction}
\label{sec:introduction}

In 2012, the ATLAS and CMS experiments observed a resonance consistent with the Higgs boson with a mass of approximately 125\GeV at the CERN LHC~\cite{ATLAS:HiggsDiscovery,CMS:HiggsDiscovery,Chatrchyan:2013lba},
providing strong evidence for spontaneous symmetry breaking
via the Brout--Englert--Higgs mechanism~\cite{Higgs:1964ia,Higgs:1964pj,Guralnik:1964eu,Higgs:1966ev,Kibble:1967sv,Englert:1964et}.
The observation was followed by precision measurements of the mass, couplings, and CP quantum numbers of the new boson,
which were found to be consistent with the predictions of the standard model (SM) of particle physics~\cite{Khachatryan:2014kca,Aad:2015zhl,Aad:2015mxa,Khachatryan:2016vau,Sirunyan:2017exp}.

Several extensions of the SM predict a more complex Higgs sector with several Higgs fields,
yielding a spectrum of Higgs bosons with different masses, charges, and other properties.
These models are constrained, but not excluded, by the measured properties of the 125\GeV boson.
The observation of additional Higgs bosons would provide unequivocal evidence for the existence of physics beyond the SM.
Two-Higgs-doublet models (2HDMs) predict five different Higgs bosons:
two neutral CP-even particles \Ph and \PH (with $m_\Ph \leq m_\PH$), one neutral CP-odd particle {\PSA}, and two charged Higgs bosons \Hpm~\cite{2HDM}.

The 2HDMs are classified into different types,
depending on the coupling of the two Higgs doublets to fermions.
This search is interpreted in the context of the
``type II" 2HDM, where one doublet couples to down-type quarks and charged leptons, and the other to up-type quarks.
The minimal supersymmetric standard model (MSSM) Higgs sector is a type II 2HDM~\cite{MSSM}.
At tree level, the Higgs sector of a type II 2HDM can be described with two parameters.
In the context of \Hpm searches, they are conventionally chosen
to be the mass of the charged Higgs boson (\mHpm) and the ratio of the vacuum expectation values of the two Higgs doublets, denoted as $\tanbeta$.
Charged Higgs bosons
are also predicted by more complex models, such as triplet models~\cite{Senjanovic:1975rk,Gunion:1989ci,Georgi:1985nv}.

\begin{figure}[hbt]
	\centering
	\includegraphics[width=0.32\textwidth]{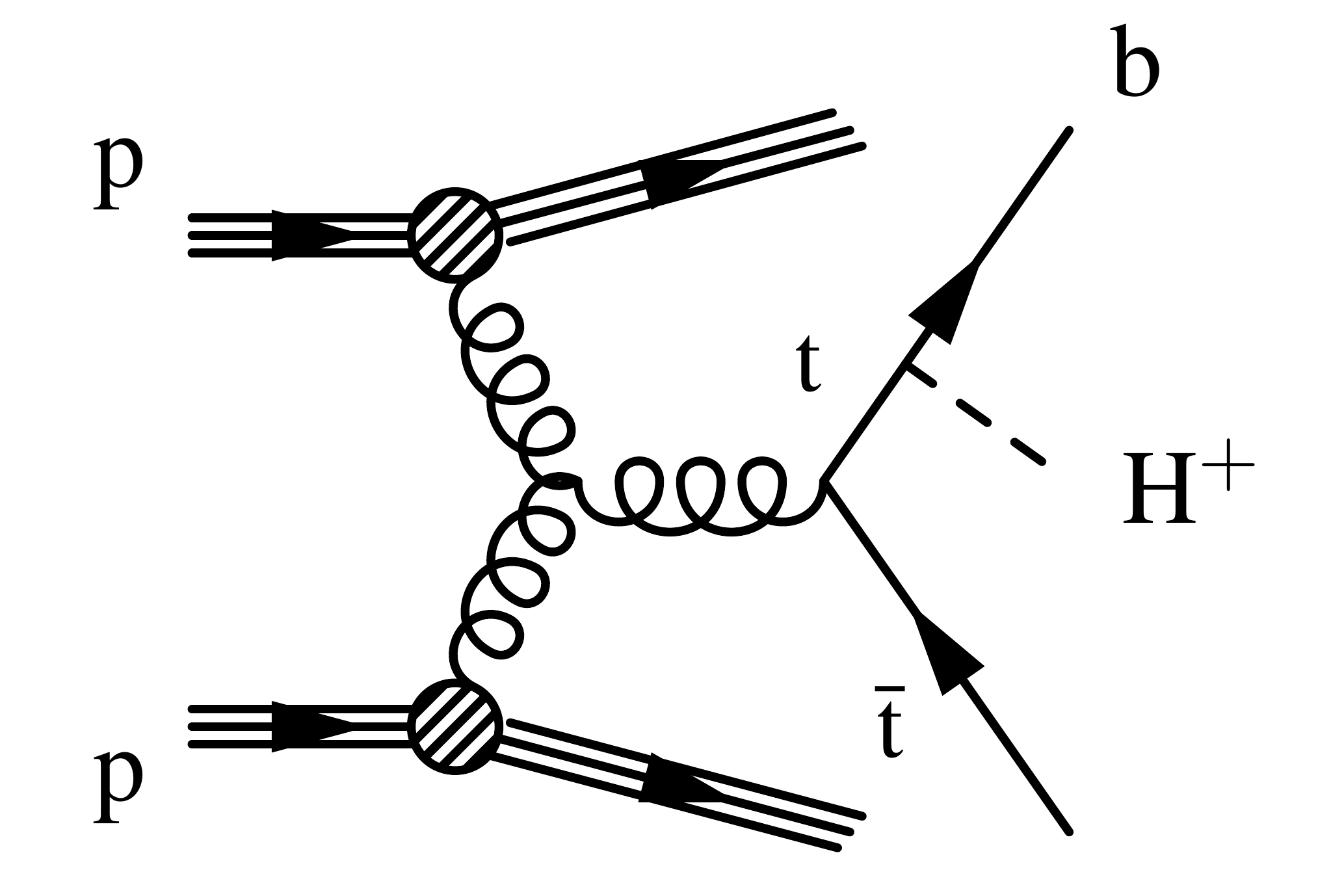}
	\includegraphics[width=0.32\textwidth]{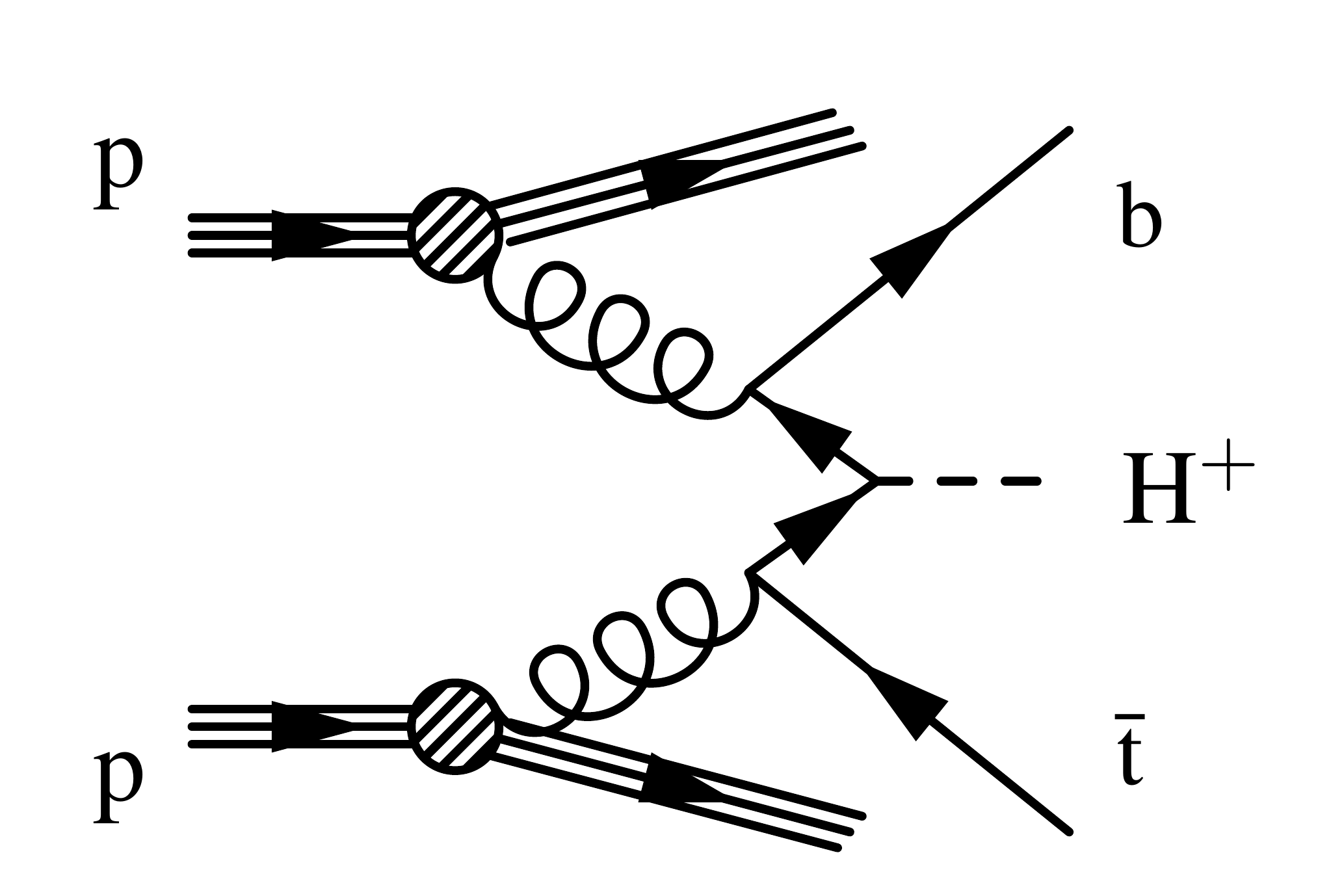}
	\includegraphics[width=0.32\textwidth]{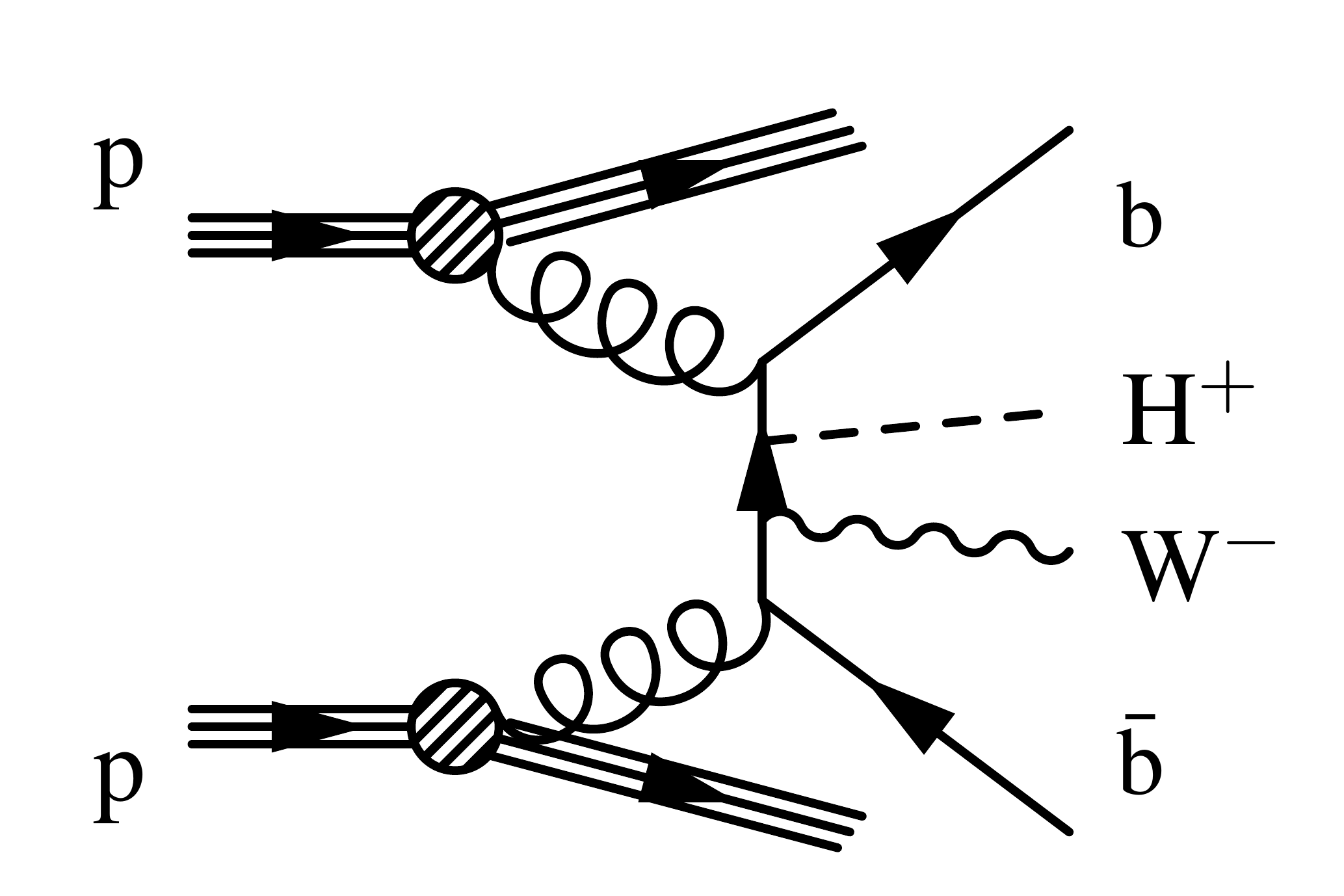}
	\caption{Leading order diagrams describing charged Higgs boson production.
	Double-resonant top quark production (left) is the dominant process for light \Hpm,
	whereas the single-resonant top quark production (middle) dominates for heavy \Hpm masses.
	For the intermediate region (\mHpmIntermediate), both production modes and their interplay with the
	nonresonant top quark production (right) must be taken into account.
	Charge-conjugate processes are implied.}
	\label{fig:feynman}
\end{figure}

The dominant production mechanism of the \Hpm depends on its mass.
Examples of leading order (LO) diagrams describing the \Hpm production in 2HDM in different mass regions are shown in Fig.~\ref{fig:feynman}.
Light \Hpm, with a mass smaller than the mass difference between the top and the bottom quarks (\mHpmLight), are predominantly produced in decays of top quarks
(double-resonant top quark production, Fig.~\ref{fig:feynman} left), whereas heavy \Hpm (\mHpmHeavy) are produced in association with a top quark as
\heavyProduction (single-resonant top quark production, Fig.~\ref{fig:feynman} middle).
In the intermediate region near the mass of the top quark (\mHpmIntermediate),
the nonresonant top quark production mode (Fig.~\ref{fig:feynman} right) also contributes and the full $\Pp\Pp\to\Hpm\PW^\mp\cPqb\cPaqb$ process must be calculated
in order to correctly account for all three production mechanisms and their interference~\cite{Degrande:2016hyf}.

In type II 2HDM, a light \Hpm decays almost exclusively to a tau lepton and a neutrino.
For the heavy \Hpm, the decay into top and bottom quarks
($\Hplus \to \cPqt\cPaqb$ and $\Hminus \to \cPaqt\cPqb$, together denoted as $\Hpm \to \cPqt\cPqb$)
is dominant, but
since the coupling of the \Hpm to leptons is proportional to $\tanbeta$,
the branching fraction to a tau lepton and a neutrino
($\Hplus \to \Pgt^{+}\Pgngt$ and $\Hminus \to \Pgt^{-}\Pagngt$, together denoted as \Hpmtaunu)
remains sizable for large values of $\tanbeta$.

Direct searches for \Hpm have been performed
at LEP~\cite{Abbiendi:2013hk}, at the Fermilab Tevatron~\cite{Aaltonen:2009vf,Abazov:2011up}, and by the LHC experiments.
The ATLAS and CMS Collaborations have covered
several \Hpm decay channels, such as \taunu~\cite{Aad:2012tj,Chatrchyan:2012vca,Aad:2012rjx,Aad:2014kga,Khachatryan:2015qxa,Aaboud:2016dig,Aaboud:2018gjj}, \cPqt\cPqb{}~\cite{Khachatryan:2015qxa,Aad:2015typ,Aaboud:2018cwk}, \cPqc\cPqs{}~\cite{Aad:2013hla,Khachatryan:2015uua}, \cPqc\cPqb{}~\cite{Sirunyan:2018dvm} and $\PW^{\pm}\cPZ$~\cite{Aad:2015nfa,Sirunyan:2017sbn},
in their previous searches at center-of-mass energies of 7, 8, or 13\TeV.
Additionally, the ATLAS and CMS results on searches
for additional neutral Higgs bosons have been interpreted in the 2HDM parameter space,
constraining the allowed \Hpm mass range as a function of $\tanbeta$~\cite{Aaboud:2017sjh,Sirunyan:2018zut,Sirunyan:2018taj,Arbey:2017gmh}.

In this paper, a direct search for \Hpm decaying into a tau lepton and a neutrino is presented, based on data collected
at a center-of-mass energy of 13\TeV by the CMS experiment in 2016, corresponding to an integrated luminosity of $35.9\fbinv$.
The search is conducted in three different final states, labeled in this paper as
the hadronic final state (\tauhjets, where \tauh denotes a hadronically decaying tau lepton), 
the leptonic final state with a \tauh (\ltauh), and the leptonic final state without a \tauh (\lnotauh). 
For the hadronic final state, events contain a \tauh, missing transverse momentum due to neutrinos,
and additional hadronic jets from top quark decays and {\cPqb} quarks.
The leptonic final state with a \tauh contains a single isolated lepton (electron or muon), missing transverse momentum, hadronic jets and a \tauh. 
The leptonic final state without a \tauh is defined in a similar way, except that events with a \tauh are rejected.
In the leptonic final states, the lepton can originate either from the decays of the tau leptons from \Hpm decays, or from a $\PW^{\pm}$ boson decay.

In each final state, events are further classified into different categories for statistical analysis.
A transverse mass distribution is reconstructed in each category of each final state and used
in a maximum likelihood fit to search for an \Hpm signal.
The \Hpm mass range from 80\GeV to 3\TeV is covered in the search,
including the intermediate mass range near \mtop.

This paper is organized as follows.
The CMS detector is briefly presented in Section \ref{sec:cmsdet}.
The methods used in event simulation and reconstruction are described in Sections \ref{sec:sim} and \ref{sec:reco}, respectively.
The event selection and categorization criteria are presented in Section \ref{sec:selection},
while Section \ref{sec:bg} details the background estimation methods used in the analysis.
Systematic uncertainties included in the analysis are described in Section \ref{sec:uncertainties}.
Finally, the results are presented in
Section \ref{sec:results}
and summarized in
Section \ref{sec:conclusion}.

\section{The CMS detector}
\label{sec:cmsdet}

The central feature of the CMS apparatus is a superconducting solenoid of 6\unit{m} internal diameter, providing a magnetic field of 3.8\unit{T}. Within the solenoid volume are a silicon pixel and strip tracker, a lead tungstate crystal electromagnetic calorimeter (ECAL), and a brass and scintillator hadron calorimeter, each composed of a barrel and two endcap sections. Forward calorimeters extend the pseudorapidity ($\eta$) coverage provided by the barrel and endcap detectors up to $\abs{\eta} = 5$.
Muons are detected in gas-ionization chambers embedded in the steel flux-return yoke outside the solenoid.
Events of interest are selected using a two-tiered trigger system~\cite{Khachatryan:2016bia}. The first level, composed of custom hardware processors, uses information from the calorimeters and muon detectors to select events at a rate of around 100\unit{kHz} within a time interval of less than 4\mus. The second level, known as the high-level trigger (HLT), consists of a farm of processors running a version of the full event reconstruction software optimized for fast processing, and reduces the event rate to around 1\unit{kHz} before data storage.
A more detailed description of the CMS detector, together with a definition of the coordinate system used and the relevant kinematic variables, can be found in Ref.~\cite{CMS-JINST}.

\section{Event simulation}
\label{sec:sim}

The signal samples for the light \Hpm mass values from 80 to 160\GeV are generated at next-to-leading order (NLO)
with the \MGvATNLO~v2.3.3~\cite{MadGraph} generator,
assuming \Hpm production via top quark decay ($\Pp\Pp\to\Hpm\PW^{\mp}\cPqb\cPaqb$).
For the heavy \Hpm mass range from 180\GeV to 3\TeV, the same approach is used except that \Hpm
production via \heavyProduction is assumed.
For the intermediate mass range from 165 to 175\GeV,
the samples are generated at LO
using the \MGvATNLO~v2.3.3
with the model described in Ref.~\cite{Degrande:2016hyf}, which is available only at LO.

The effect of using LO instead of NLO samples is estimated by comparing
kinematic distributions and final event yields
from both types of samples in mass regions below (150--160\GeV) and above (180--220\GeV)
the intermediate range.
Significant differences are observed in some kinematic variables such as jet multiplicity,
affecting the selection efficiency and the predicted final signal yield.
Since the shapes of the final \mT distributions are found to be compatible between the LO and the NLO samples,
a LO-to-NLO correction is performed by scaling the final signal event yield from each intermediate-mass sample.
The overall effect of the correction is to scale down the signal event yield,
resulting in more conservative results than would be obtained using LO samples without this correction.

The NLO/LO signal yield ratios are similar for all mass points within the 150--160\GeV and 180--200\GeV mass regions,
but different between these two regions.
Thus the correction factor for each final state and event category
is calculated as an average over the NLO/LO ratios of the final event yields.
This is done separately for the 150--160\GeV and 180--200\GeV regions,
and the correction derived in the 150--160\GeV region is applied to the intermediate signal sample with $\mHpm = 165\GeV$,
for which \mHpmLight and the \Hpm production is still dominated by top quark decays,
while the correction derived in the 180--200\GeV region is applied to the 170 and 175\GeV samples with \mHpmHeavy.
For all signal samples up to $\mHpm = 500\GeV$, {\scshape MadSpin}~\cite{Artoisenet:2012st} is used to model the \Hpm decay,
while \PYTHIA~8.212 is used above 500\GeV.

In the leptonic final states, where accurate modeling of jet multiplicity is needed for the correct categorization of events,
the {\scshape MG5}{\_aMC@NLO}~v2.2.2 generator~\cite{MadGraph} is used to simulate the \ttbar events at NLO.
In the hadronic final state, the statistical uncertainty in the final event yield needs to be minimized
for reliable modeling of the \mT shape of the \ttbar background,
and thus a larger sample generated using \POWHEG~v2.0~\cite{POWHEG1,POWHEG2,POWHEG3,POWHEG-TT,Frixione:2007nw}
with FxFx jet matching and merging~\cite{Frederix:2012ps} is used to model this background.
The \POWHEG v2.0 generator is used to model single top quark production via $t$-channel and \cPqt\PW~production~\cite{POWHEG-ST,Re:2010bp},
while the \MGvATNLO~v2.2.2 generator is used for the $s$-channel production.
The value of \mtop is set to 172.5\GeV for all \ttbar and single top quark samples.
The \PW+jets and \DY events are generated at LO using \MGvATNLO~v2.2.2 with up to four noncollinear partons in the final state~\cite{Alwall:2007fs}.
The diboson processes ({\PW}{\PW}, {\PW}{\cPZ}, {\cPZ}{\cPZ}) are simulated using \PYTHIA~8.212.

The simulated samples are normalized to the theoretical cross sections
for the corresponding processes.
For the \ttbar background and the single top quark background in the $s$ and \cPqt\PW{} channels,
the cross sections are calculated
at next-to-NLO precision~\cite{Czakon:2011xx,Kidonakis:2013zqa}.
NLO precision calculations are used
for single top quark production in the $t$ channel, and for the \PW+jets, \DY,
and diboson processes~\cite{Kant:2014oha,Kidonakis:2013zqa,Melnikov:2006kv,Campbell:2011bn}.

For all simulated samples, the NNPDF3.0 parton distribution functions (PDFs)~\cite{NNPDF3} are used, and the generators are interfaced with \PYTHIA~8.212 to model the parton showering, fragmentation, and the decay of the tau leptons.
The \PYTHIA~parameters affecting the description of the underlying event are set to the CUETP8M1 tune~\cite{TuneCUETP8M1} for all processes
except \ttbar, for which a customized CUETP8M2T4 tune~\cite{CMS:2016kle} is used.

Generated events are processed through a simulation of the CMS detector based on
the \GEANTfour v9.4 software~\cite{GEANT4}, and they are reconstructed following the same algorithms that are used for data.
The effect of additional soft inelastic proton-proton ($\Pp\Pp$) interactions (pileup) is modeled by generating minimum bias collision events with \PYTHIA and mixing them with the simulated hard scattering events.
The effects from multiple inelastic $\Pp\Pp$ collisions occurring per bunch crossing (in-time pileup), as well as the effect of
inelastic collisions happening in the preceding and subsequent bunch crossings (out-of-time pileup) are taken into account.
The simulated events are weighted such that the final pileup distribution matches the one observed in data.
For the data collected in 2016, an average of approximately 23 interactions per bunch crossing was measured.

\section{Event reconstruction}
\label{sec:reco}

Event reconstruction is based on the particle-flow (PF) algorithm~\cite{CMS-PRF-14-001} that aims to reconstruct and identify each individual particle in an event with an optimized combination of information from the various elements of the CMS detector.
The output of the PF algorithm is a set of PF candidates, classified into
muons, electrons, photons, and charged and neutral hadrons.

The collision vertices are reconstructed from particle tracks using the deterministic annealing algorithm~\cite{VTX}.
The reconstructed vertex with the largest value of the physics-object transverse momentum squared
($\pt^2$) sum is taken to be the primary {\Pp\Pp} interaction vertex.
The physics objects in this case are the jets, clustered using the anti-\kt jet finding algorithm~\cite{Cacciari:2008gp,Cacciari:2011ma} with the tracks assigned to the vertex as inputs, and the associated missing transverse momentum, calculated as the negative vector sum of the \pt of those jets.
All other reconstructed vertices are attributed to pileup.

Electrons are reconstructed and their momentum is estimated by combining the momentum measurement from the tracker at the interaction vertex with the energy measurement in the ECAL. The energy of the corresponding ECAL cluster and the energy sum of all bremsstrahlung photons spatially compatible with originating from the electron tracks are taken into account.
The momentum resolution for electrons with $\pt\approx45\GeV$ from $\cPZ \to \Pe\Pe$ decays ranges from 1.7\% for nonshowering electrons in the barrel region to 4.5\% for showering electrons in the endcaps~\cite{Khachatryan:2015hwa}.
In addition, electrons are required to pass an identification requirement based on a multivariate discriminant that combines several variables describing the shape of the energy deposits in the ECAL, as well as the direction and quality of the associated tracks~\cite{Hocker:2007ht}.
A tight working point with 88\% identification efficiency for \ttbar events is used to select events with an electron,
while a loose working point with 95\% efficiency is used to veto events with one or several electrons, depending on the final state.

Muons are identified as tracks in the central tracker, consistent with either a track or several hits in the muon chambers,
and associated with calorimeter deposits compatible with the muon hypothesis~\cite{Sirunyan:2018fpa}.
The momenta of muons are obtained from the curvatures of the corresponding tracks.
Contributions from other particles misidentified as muons are suppressed with a discriminant based on the track fit quality.
Two working points as defined in Ref.~\cite{Sirunyan:2018fpa} are used:
a medium working point with 97\% identification efficiency is used to select events with a muon,
while a loose working point with ${>}99\%$ identification efficiency is used for vetoing muons.

The background contributions from nonprompt and misidentified leptons are
suppressed by requiring the leptons to be isolated from hadronic activity in the event.
For this purpose, an isolation discriminant is defined as the \pt sum of the PF candidates in a cone around the lepton,
divided by the \pt of the lepton.
For optimal performance across the lepton momentum range, the cone size is varied with the lepton \pt as
$\Delta{R} = \sqrt{\smash[b]{(\Delta\eta)^2+(\Delta\phi)^2}} = 10\GeV / \text{min}(\text{max}(\pt, 50\GeV),\allowbreak{200\GeV)}$,
where $\Delta\phi$ denotes a difference in azimuthal angle,
leading to cone radii from 0.05 to 0.20.
A tight (loose) isolation criterion with discriminant $<0.1$ ($0.4$) is used in lepton selection (veto).

For each event, hadronic jets are clustered from the reconstructed PF candidates using the infrared and collinear safe anti-\kt algorithm~\cite{Cacciari:2008gp, Cacciari:2011ma} with a distance parameter of 0.4. The jet momentum is determined as the vectorial sum of all particle momenta in the jet, and is found from simulation to be within 5 to 10\% of the true momentum over the whole \pt spectrum and detector acceptance.
Pileup can contribute additional tracks and calorimetric energy deposits to the jet momentum. To mitigate this effect, tracks identified as originating from pileup vertices are discarded and an offset correction is applied to correct for remaining contributions. Jet energy corrections are derived from simulation to bring the measured response of jets to that of particle level jets on average. In situ measurements of the momentum balance in dijet, $\text{photon} + \text{jet}$, $\cPZ + \text{jet}$, and multijet events are used to account for any residual differences in jet energy scale between data and simulation~\cite{Khachatryan:2016kdb}. The jet energy resolution amounts typically to 15\% at 10\GeV, 8\% at 100\GeV, and 4\% at 1\TeV~\cite{CMS:2017wyc}. Additional selection criteria are applied to each jet to remove jets potentially dominated by anomalous contributions from various subdetector components or reconstruction failures.

Jets originating from the hadronization of {\cPqb} quarks ({\cPqb} jets)
are identified using the combined secondary vertex algorithm~\cite{btag:csv, Sirunyan:2017ezt},
which uses information on the decay vertices of long-lived hadrons and the
impact parameters of charged particle tracks as input to a neural network discriminant.
The working point is chosen such that the probability to misidentify jets originating from light-flavor quarks or gluons ({\cPqc} quarks) as {\cPqb} jets is 1\% (12\%),
corresponding to 63\% efficiency for the selection of genuine {\cPqb} jets in \ttbar events.
Simulated samples are corrected for differences in {\cPqb} jet identification and misidentification efficiency compared to the data.

The \tauh are reconstructed with the hadron-plus-strips algorithm~\cite{HPS,Sirunyan:2018pgf},
which uses clustered anti-\kt jets as seeds.
The hadron-plus-strips algorithm reconstructs different {\Pgt} decay modes with one charged pion and up to two neutral pions (one-prong), or three charged pions (three-prong).
Since neutral pions decay promptly to a photon pair, they are reconstructed by defining strips
of ECAL energy deposits in the $\eta$--$\phi$ plane.
The \tauh candidates are rejected if they are consistent with the hypothesis of being muons or electrons misidentified as \tauh.
The jets originating from the hadronization of quarks or gluons misidentified as \tauh are suppressed using a multivariate discriminant~\cite{Sirunyan:2018pgf}.
It combines information on \tauh isolation, based on the surrounding hadronic activity, and on its lifetime, inferred from the tracks of the \tauh decay products.
A loose working point is used for this discriminant, corresponding to ${\approx}50\%$ identification efficiency, determined from $\cPZ/\gamma^{*}\to\Pgt^{+} \Pgt^{-}$ events,
and $3\times10^{-3}$ probability for misidentifying a jet as a \tauh, determined from quantum chromodynamics (QCD) multijet events.
A correction to the energy scale
is derived using $\Pe\tauh$ and $\Pgm\tauh$ final states of
$\cPZ/\gamma^{*} \to \Pgt^{+} \Pgt^{-}$ events~\cite{Sirunyan:2018pgf}
and applied in simulated samples.

The missing transverse momentum (\ptvecmiss) is defined as the negative vector sum of the \pt of all reconstructed PF candidates~\cite{CMS-PAS-JME-17-001}.
The energy scale corrections applied to jets and \tauh are propagated to the \ptvecmiss.

The transverse mass is defined as
\begin{equation}
  \label{eq:mt}
  \mT(\tauh/\ell) = \sqrt{2 \pt(\tauh/\ell) \ptmiss (1-\cos \Delta\phi(\ptvec(\tauh/\ell),\ptvecmiss))},
\end{equation}
where $\ell$ is a generic symbol used to label the electron or muon present in the leptonic final states,
while the leading \tauh is used in the \mT in the hadronic final state.

\section{Event selection}
\label{sec:selection}

The search is conducted in three exclusive final states:
\begin{itemize}
    \item \tauhjets: hadronic final state (events with an electron or a muon are vetoed);
    \item \ltauh: leptonic final state with a hadronically decaying tau lepton (events with additional electrons or muons are vetoed); and
    \item \lnotauh: leptonic final state without a hadronically decaying tau lepton (events with a \tauh or additional electrons or muons are vetoed).
\end{itemize}

In the low-\mHpm region, below \mtop,
the sensitivity of the hadronic final state is limited by the relatively high trigger thresholds,
making the leptonic final states most sensitive for the \Hpm signal.
In the high-\mHpm region, above \mtop, the hadronic final state dominates the sensitivity,
since the selection efficiency is higher as a result of more inclusive jet multiplicity requirements.

The event selection and categorization strategies are chosen separately for each final state
to efficiently discriminate against the background events,
while ensuring a sufficient signal selection efficiency.

\subsection{Hadronic final state \texorpdfstring{(\tauhjets)}{}}

An HLT algorithm requiring the presence of a \tauh candidate and
trigger-level missing transverse momentum estimated from calorimeter information (\ptmisscalo)
is used to select the events for offline analysis.
The trigger requires the \tauh candidate to be loosely isolated with $\pt > 50\GeV$ and $\abs{\eta}<2.1$,
and with a leading track transverse momentum $\ptldgtrk > 30\GeV$.
The \ptmisscalo is required to be larger than 90\GeV.

The trigger efficiencies for the \tauh and \ptmisscalo requirements are measured separately.
The efficiency of the \tauh part of the trigger is determined with
the tag-and-probe technique~\cite{Chatrchyan:2012xi}, using $\cPZ/\gamma^{*}\to\Pgt^{+} \Pgt^{-}$ events
with one hadronic and one muonic tau lepton decay.
The efficiency is found to vary between 50 and 100\%, as a function of \pt and $\eta$ of the \tauh.
The efficiency of the \ptmisscalo part of the trigger is measured
from events with a signal-like topology selected with a single-\tauh trigger,
resulting in efficiencies between 10 and 100\%, depending on the value of the \ptmiss.
The simulated events are corrected to match the trigger efficiencies measured in the data.

In the offline selection, low thresholds for the \pt of the reconstructed \tauh and \ptmiss are needed to maximize the sensitivity for light \Hpm.
Thus selection criteria identical to those in the HLT are applied to the reconstructed \tauh
candidate and to the \ptmiss.
The one-prong \tauh candidates, corresponding to \Pgt{} decays into a charged pion and up to two neutral pions,
are selected for further analysis.
Events are required to contain at least three jets with $\pt>30\GeV$ and  $\abs{\eta}<4.7$,
separated from the reconstructed \tauh by ${\Delta{R}>0.5}$.
At least one of the jets is required to pass the {\cPqb} jet identification with $\abs{\eta}<2.4$.
Any event with isolated electrons (muons) with $\pt>15 (10)\GeV$, $\abs{\eta}<2.5$, and passing the loose identification and isolation criteria is rejected.

\begin{figure}[tb]
	\centering
	\includegraphics[width=0.49\textwidth]{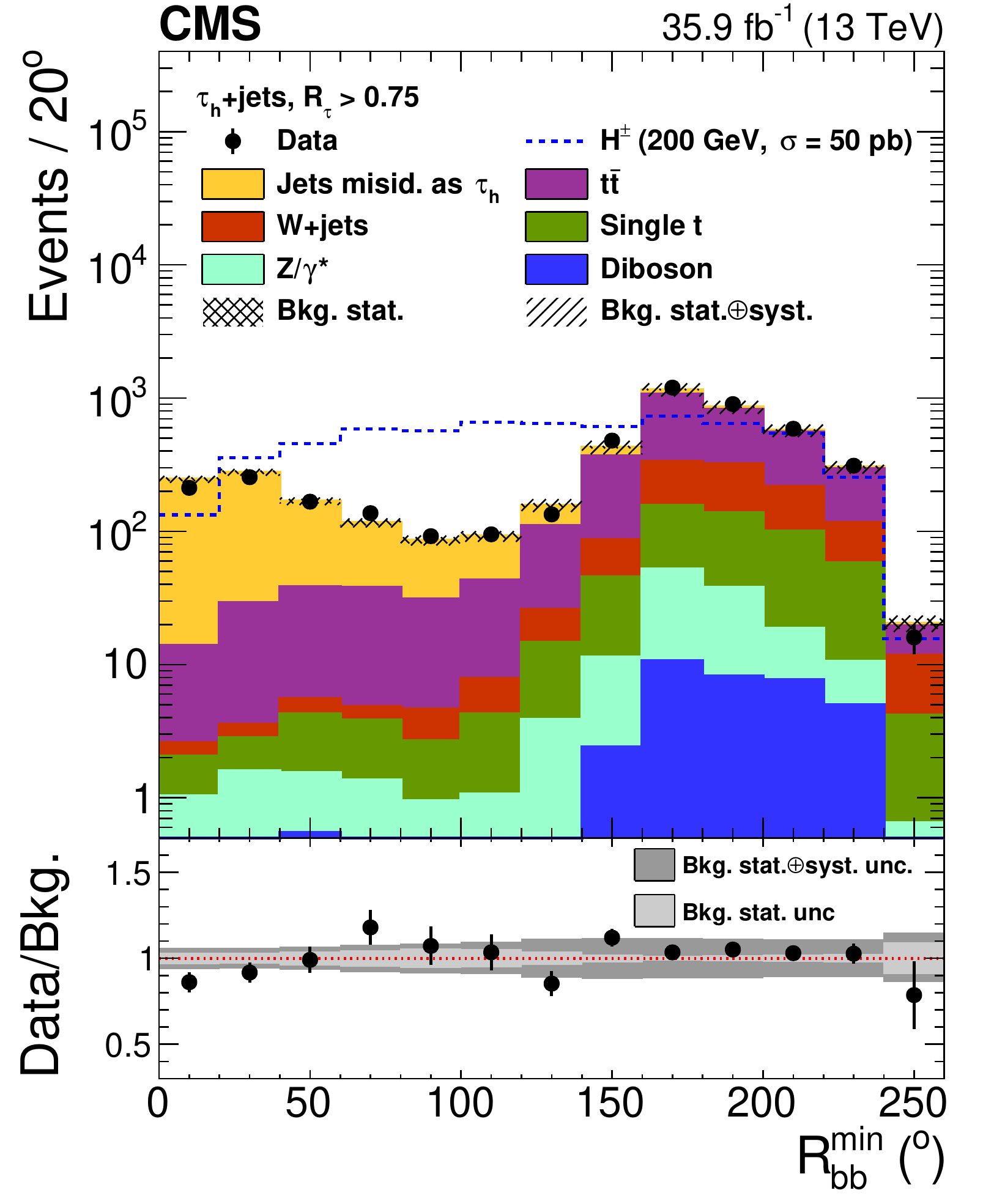}
	\includegraphics[width=0.49\textwidth]{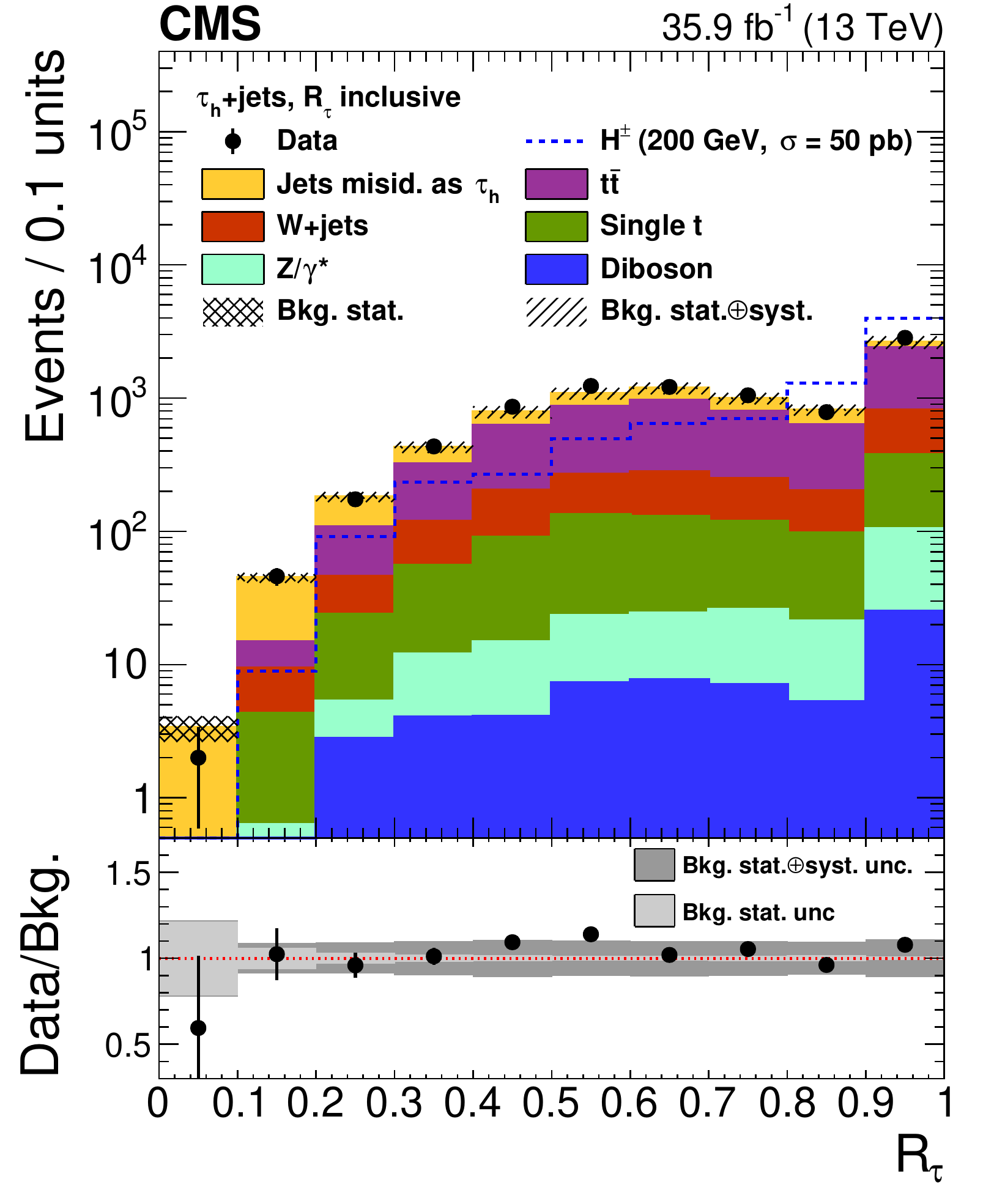}
	\caption{
       The distribution of the angular discriminant $\Rbbmin$ after all other selections including the $\Rtau=\Rtaudef > 0.75$ requirement have been applied (left),
       and the distribution of the $\Rtau$ variable used for categorization after all other selections including the $\Rbbmin > 40^{\circ}$ requirement have been applied (right).
       	}
	\label{fig:RbbRtau}
\end{figure}

To suppress the background from QCD multijet events
with a jet misidentified as a \tauh,
an additional selection based on $\Delta\phi(\tauh,\ptmiss)$ and $\Delta\phi(\text{jet}_n,\ptmiss)$ is applied, where
the index $n$ runs over the three highest \pt jets ($\text{jet}_n$) in the event.
QCD multijet events passing the previous selection steps
typically contain a hadronic jet misidentified as a \tauh, another hadronic jet recoiling in the opposite direction,
and \ptvecmiss arising from the mismeasurement of the jet momenta.
These events can be suppressed with an angular discriminant defined as
\begin{equation}
  \label{eq:rbbmin}
  \Rbbmin = \min_n\left\{
  \sqrt{\left(180^\circ - \Delta\phi(\tauh,\ptvecmiss)\right)^2 + \left(\Delta\phi(\text{jet}_n,\ptvecmiss)\right)^2 } \right\}.
\end{equation}
The selected events are required to have $\Rbbmin > 40^{\circ}$.
The distribution of the $\Rbbmin$ variable after all other selections is shown in Fig. \ref{fig:RbbRtau} (left).

The selected events are classified into two categories based on the value of the variable
$\Rtau=\Rtaudef$, reflecting the helicity correlations emerging from the opposite polarization states of the tau leptons originating from $\PW^{\pm}$ and \Hpm decays~\cite{Roy:1999xw}.
The distribution of the $\Rtau$ variable is shown in Fig. \ref{fig:RbbRtau} (right).
After all other selections, most of the signal events have a large value of \Rtau, and
the high-\Rtau category provides a good signal-to-background ratio.
For large \mHpm values, the signal events are more evenly distributed between the two categories,
so inclusion of the background-dominated low-\Rtau category in the statistical analysis further improves the sensitivity for the heavy \Hpm.
Separating the two categories at $\Rtau = 0.75$ maximizes the signal sensitivity across the \mHpm range.

\subsection{Leptonic final state with a hadronically decaying tau lepton \texorpdfstring{(\ltauh)}{}}
\label{sec:evsel:ltauh}

Single-lepton trigger algorithms are used for the online selection of events with isolated electrons or muons.
Several HLT algorithms for electron (muon) selection with different thresholds starting from 27 (24)\GeV, with $\abs{\eta}<2.1$ ($2.4$)
and with different isolation criteria,
are used in \textit{or} combination to maximize the efficiency across the lepton \pt range.

In the offline selection, electrons (muons) are required to have $\pt>35 (30)\GeV$ and $\abs{\eta}<2.1 (2.4)$
because of trigger constraints. Electrons (muons) are required to pass the tight (medium) identification and tight isolation requirements.
Events with any additional electrons (muons) with $\pt>10\GeV$ and $\abs{\eta}<2.1 (2.4)$ that pass the loose identification and isolation criteria are vetoed.
Efficiencies for online and offline identification of leptons are measured, and the simulated events are corrected to match the efficiencies observed in data.
The presence of a \tauh is required, with
$\pt>20\GeV$, $\abs{\eta}<2.3$, and with a $\Delta{R}$ separation of at least 0.5 with respect to the lepton.

One, two, or three jets are required with $\pt>30\GeV$ and  $\abs{\eta}<2.4$, separated from the lepton and the \tauh by $\Delta{R}>0.5$.
At least one of the jets is required to pass the {\cPqb} jet identification.
To suppress the background from jets misidentified as \tauh, the \ptmiss is required to be at least $70\GeV$.
The background contribution from events with muons originating from {\cPqb} hadron decays is suppressed by
requiring $\Delta\phi(\ell, \ptvecmiss)$ to exceed 0.5.

The selected events are classified into several categories for statistical analysis.
Three categories are defined based on the jet multiplicity and the number of jets passing the {\cPqb} jet identification: 1j1b (one jet that is also identified as a {\cPqb} jet), $\geq$2j1b, and $\geq$2j$\geq$2b.
A second categorization is performed in bins of \ptmiss: 70--100, 100--150, and ${>}150\GeV$.
Together with the separate electron and muon final states, this results in 18 categories.

The signal-to-background ratio in different categories varies with \Hpm mass, as jet categories with two jets and high \ptmiss become more sensitive for higher \mHpm values.
The background-enriched categories allow a precise determination of the background yields with a fit to data
and extrapolation of this information to signal regions.
The categorization is found to improve the expected sensitivity significantly,
especially in the low-\mHpm region, where efficient discrimination against backgrounds is essential.

\subsection{Leptonic final state without a hadronically decaying tau lepton \texorpdfstring{(\lnotauh)}{}}
\label{sec:evsel:lnotauh}

The event selection criteria for the \lnotauh final state are identical to those described in Section \ref{sec:evsel:ltauh} for the \ltauh final state,
except for the following requirements. An event is vetoed if it contains a \tauh with $\pt>20\GeV$, $\abs{\eta}<2.3$, and with a $\Delta{R}$ separation of at least 0.5 with respect to the lepton. Two or three jets are required, each jet separated from the lepton by $\Delta{R}>0.5$. Higher jet multiplicities are not selected, because they are
expected to be more sensitive in searches for other \Hpm decay modes, such as $\Hpm\to\cPqt\cPqb$.
At least one of the jets is required to pass the {\cPqb} jet identification.

The number of QCD multijet events with jets misidentified as leptons is reduced to a negligible level
by requiring a high \ptmiss of ${>}100\GeV$ and by applying the following angular selections:
\begin{itemize}
    \item $\Delta\phi(\ell, \ptvecmiss)>0.5$;
    \item $\Delta\phi(\text{leading jet}, \ptvecmiss)>0.5$; and
    \item $\min(\Delta\phi(\ell, \text{jet}_n)) < \pi - 0.5$,
\end{itemize}
where $\text{jet}_n$ refers to any of the selected jets in the events.
The first criterion is identical to the one applied in the \ltauh final state against muons from {\cPqb} hadron decays whereas
the second discriminates efficiently against the QCD multijet background.
The last requirement is designed to reject background events where
all the jets are back-to-back with respect to the selected lepton.

To further enhance the signal sensitivity and to constrain the backgrounds, a similar categorization as in the \ltauh final state is established. Four categories are used based on jet multiplicity and the number of jets passing the {\cPqb} jet identification: 2j1b, 2j2b, 3j1b, and 3j$\geq$2b, followed by two categories in \ptmiss:
100--150 and ${>}150\GeV$. Together with the separate electron and muon final states, this results in 16 categories.

An overview of the event selection criteria in all three final states is shown in Table~\ref{tab:selections}.

\begin{table}
	\centering
	\topcaption{A summary of the event selection criteria applied in each final state. The electrons, muons, \tauh candidates and jets are required to be separated from each other by $\Delta R > 0.5$ in all final states. The $\dag$ symbol means that the selection is identical between \ltauh and \lnotauh final states. In all final states, events with additional electrons or muons are vetoed as detailed in Section~\ref{sec:selection}. In this table, ``b jets" refers to all jets passing the b jet identification, and $\text{jet}_n$ refers to any of the selected jets.}
	\label{tab:selections}
\renewcommand{\arraystretch}{1.1}
  \cmsTable{
  \begin{tabular}{ l l l l }
  Selection & \tauhjets & \ltauh & \lnotauh \\
  \hline
  Trigger                       & \tauh+\ptmisscalo & single \Pe~or single \Pgm & \dag \\
  [\cmsTabSkip]
  Number of \tauh candidates     & $\geq1$     & $\geq1$ & $0$ \\
  \tauh \pt & $\pt > 50\GeV$, $\ptldgtrk > 30\GeV$ & $\pt>20\GeV$ & \NA \\
  \tauh $\abs{\eta}$ & $\abs{\eta}<2.1$ & $\abs{\eta}<2.3$ & \NA \\
  [\cmsTabSkip]
  Number of electrons and muons     & $0$      & 1\Pe~or 1 \Pgm~(exclusively) & \dag \\
  Electron  \pt               & \NA & $\pt>35\GeV$ & \dag \\
  Ekectron  $\abs{\eta}$      & \NA & $\abs{\eta}<2.1$ & \dag \\
  Muon \pt                  & \NA & $\pt>30\GeV$ & \dag \\
  Muon $\abs{\eta}$                  & \NA & $\abs{\eta}<2.4$ & \dag \\
  [\cmsTabSkip]
  Number of jets (incl. b jets)    & $\geq3$ jets & 1--3 jets & 2--3 jets \\
  Jet \pt                 & $\pt>30\GeV$ & $\pt>30\GeV$ & \dag \\
  Jet $\abs{\eta}$                & $\abs{\eta}<4.7$ & $\abs{\eta}<2.4$ & \dag \\
  [\cmsTabSkip]
  Number of b jets & $\geq1$ b jets & 1--3 b jets & \dag \\
  b jet $\abs{\eta}$           & $\abs{\eta}<2.4$ & $\abs{\eta}<2.4$ & \dag \\
  [\cmsTabSkip]
  \ptmiss                       & $\ptmiss > 90\GeV$ & $\ptmiss > 70\GeV$ & $\ptmiss > 100\GeV$ \\
  [\cmsTabSkip]
  Angular selections & $\Rbbmin > 40^{\circ}$ &  $\Delta\phi(\ell, \ptmiss)>0.5$ &  $\Delta\phi(\ell, \ptvecmiss)>0.5$, \\ 
                                      & & ($\ell = \Pe$ or \Pgm) & $\Delta\phi(\text{leading jet}, \ptvecmiss)>0.5$, \\
                                      & & & $\min(\Delta\phi(\ell, \text{jet}_n)) < \pi - 0.5$ \\
  \hline
  \end{tabular}
  }   
\renewcommand{\arraystretch}{1.0}
\end{table}

\section{Background estimation}
\label{sec:bg}

The dominant background processes in the hadronic final state are QCD multijet and \ttbar production. Other backgrounds are single top quark production, \PW{} boson production in association with jets, \DY processes, and diboson production.
We refer to \ttbar and single top quark events as ``top events", and to \PW+jets, \DY, and diboson events as ``electroweak events".
The backgrounds from events containing either a genuine \tauh or an electron or a muon misidentified as a \tauh are estimated
from simulation, while the background from jets misidentified as a \tauh is estimated from data.
The correct identification or misidentification of a \tauh is determined
by requiring a generator-level tau lepton to match with the reconstructed \tauh within a $\Delta{R}$ cone of 0.1.

In the events where a jet is misidentified as a \tauh (denoted as \jetToTauh), QCD multijet production is the dominant process.
The jet $\to$ \tauh background is estimated using a control sample enriched in jets misidentified as \tauh,
obtained by inverting the offline \tauh isolation requirement used for signal selection.
The contamination of the control region from electroweak/top events with a genuine \tauh
or a lepton misidentified as a \tauh
is estimated from the simulation and subtracted from the control sample.
The difference in selection efficiency between signal and control regions
is corrected by normalizing the control sample with \textit{fake factors},
calculated at an early stage of event selection
(\ie before applying {\cPqb} jet identification, offline selection on \ptmiss or the angular selections),
where a possible signal does not stand out from the large background yield.
To account for the correlation between the \pt of the \tauh and \ptmiss as well as
geometrical differences in detector response,
the measurement is performed in bins of \pt and $\abs{\eta}$ of the \tauh.

The \jetToTauh background consists of two components: the QCD multijet events
and electroweak/top events with jets misidentified as \tauh.
The jets in these two background components have different quark and gluon composition implying different tau fake factors.
Thus the fake factors for misidentified \tauh from the QCD multijet events and for
misidentified \tauh from electroweak/top events are estimated separately.
The fake factor for the QCD multijet events is defined as the ratio of the QCD multijet event yields in signal and control regions.
The QCD multijet event yield in the control region is estimated
by subtracting the simulated electroweak/top contribution (both genuine and non-genuine \tauh events) from data.
To estimate the contribution of the QCD multijet events in the signal region,
a binned maximum likelihood fit of \ptmiss templates to data is performed,
using the fraction of the QCD multijet events as a fit parameter.
The templates describe the expected shape of the \ptmiss distribution for each background component prior to the fit.
The \ptmiss shape of the QCD multijet events is assumed to be similar in the signal and control regions,
so the shape observed in the control region is used as the fit template.
The template for electroweak/top events is obtained directly from simulation.
The fake factor for electroweak/top events is also estimated
from simulation as the ratio of event yields in signal and control regions.
Finally, the overall normalization factor of the control sample (as a function of the \pt and $\abs{\eta}$ of the \tauh)
is determined as a weighted sum of the two fake factors, where
the weight corresponds to the relative fractions of the QCD multijet and electroweak/top events in the control region after all selections.
A closure test is performed by comparing the background predictions
obtained with the above method to data in a signal-depleted validation region.
The validation region is defined similarly to the signal region,
except that events with jets passing the {\cPqb} jet identification are vetoed.

In the leptonic final states, the dominant background is \ttbar production in which the semileptonic \ttbar decays are dominant in the \lnotauh final state and the dilepton \ttbar decays are dominant in the \ltauh final state. Minor backgrounds include single top quark, \PW+jets, \DY, and diboson production.
The QCD multijet background is suppressed to a negligible level with tight angular selections and \ptmiss requirements.
All backgrounds in the two leptonic final states are estimated from simulation.

\section{Systematic uncertainties}
\label{sec:uncertainties}

A summary of uncertainties incorporated in the analysis is given in Table~\ref{tab:systematics},
where the effects of the different uncertainties on the final event yields are shown. 
For the uncertainties common to all final states, the variations in the yields are similar across the final states. 
Some of them affect only the final event yield for a given signal or background process,
whereas others also modify the shape of the final \mT distributions.
The uncertainties from different sources are assumed to be uncorrelated.
Each uncertainty is treated as 100\% correlated among the signal and background processes,
except for the few special cases mentioned in the following.

\begin{table}
	\centering
	\topcaption{Effect of systematic uncertainties on the final event yields in per cent, prior to the fit, summed over all final states and categories. For the \Hpm signal, the values for $\mHpm = 200\GeV$ are shown.}
	\label{tab:systematics}
\renewcommand{\arraystretch}{1.1}
  \cmsTable{
  \begin{tabular}{ l c c c c c c }
  Source  & Shape & \Hpm (200\GeV)  & Jets $\to$ \tauh & \ttbar & Single \cPqt & Electroweak \\
  \hline
$\tauh+\ptmiss$ trigger efficiency & $\checkmark$ & 1.4 & 2.0 & 0.2 & 0.2 & 0.2 \\
\tauh identification & $\checkmark$ & 1.8 & 0.6 & 1.1 & 1.0 & 0.9 \\
Lepton selection efficiency & & 2.3 & \NA & 2.7 & 2.7 & 2.7 \\
Jet energy scale and resolution & $\checkmark$ & 4.7 & 0.4 & 5.1 & 9.2 & 13.4 \\
\tauh energy scale & $\checkmark$ & 0.2 & 0.6 & $<$ 0.1 & $<$ 0.1 & $<$ 0.1 \\
Unclustered \ptmiss energy scale & $\checkmark$ & 2.6 & $<$ 0.1 & 3.2 & 5.2 & 7.2 \\
{\cPqb} jet identification & $\checkmark$ & 3.6 & 0.8 & 3.1 & 3.4 & 13.8 \\
Integrated luminosity & & 2.5 & 0.4 & 2.5 & 2.5 & 2.5 \\
Pileup & $\checkmark$ & 1.1 & $<$ 0.1 & 0.8 & 1.2 & 4.0 \\
Jets misid. as \tauh estimation & $\checkmark$ & \NA & 6.1 & \NA & \NA & \NA \\
Cross section (scales, PDF) & & \NA & 0.8 & 5.5 & 5.3 & 3.6 \\
Top quark mass & & \NA & 0.4 & 2.7 & 2.2 & \NA \\
Acceptance (scales, PDF) & & 5.1 & 0.5 & 2.8 & 2.8 & 6.8 \\
\ttbar parton showering & & \NA & \NA & 6.1 & \NA &\NA  \\
[\cmsTabSkip]
Total & & 9.4 & 6.6 & 12.1 & 13.5 & 22.7 \\
  \hline
  \end{tabular}
  }
\renewcommand{\arraystretch}{1.0}
\end{table}

The simulated events are corrected to match the online and offline selection efficiencies measured in data.
For the trigger used in the \tauhjets final state,
the correction depends on the \pt of the \tauh and \ptmiss, so the corresponding uncertainty is taken into account as a shape uncertainty.

In the \ltauh and \lnotauh final states, the online selection with single-lepton triggers is incorporated into the overall lepton selection efficiency and
the corresponding normalization uncertainty.

The systematic uncertainties in identification and isolation efficiencies for \tauh, electron, and muon candidates are taken into account.
The agreement of the \tauh identification efficiency between data and simulated samples is measured using the tag-and-probe technique~\cite{Sirunyan:2018pgf}.
The uncertainty in the measurement is 5\%.
It is incorporated as a normalization uncertainty for all events with genuine tau leptons,
and anticorrelated between the \lnotauh final state and the final states with a \tauh{}.
For the \tauh with large \pt, an additional uncertainty of
$^{+5}_{-35}\%\pt/\TeV$ is applied in the hadronic final state as a shape uncertainty
to account for possible differences arising in the extrapolation of the measured efficiencies to the high-\pt range.
Simulated events with an electron or a muon misidentified as a \tauh
are weighted to obtain the misidentification rates measured in data.
The corrections are applied as a function of $\eta$
and the corresponding uncertainties are propagated to \mT distributions and incorporated as shape uncertainties.

For the selection of electrons (muons), the combined uncertainty in online selection and offline identification is 3 (4)\%.
For leptons vetoed with loose identification and isolation criteria the effect of this uncertainty in the final event yield is typically only $0.3\%$.
Both effects are included as normalization uncertainties.

The systematic uncertainties related to the calibration of energy measurement for jets, \tauh and \ptmiss are considered as shape uncertainties.
The uncertainties in the jet energy scale and jet energy resolution are
specified as a function of jet \pt and $\eta$.
The uncertainty in the \tauh energy scale is
$\pm1.2\%$ for $\pt< 400\GeV$ and $\pm3\%$ otherwise~\cite{Sirunyan:2018pgf}.
The variations of the jet and \tauh energy scales are propagated to \ptvecmiss,
for which the uncertainties arising from the unclustered energy deposits in the detector are also included.
The uncertainty in the lepton energy scale is negligible for this analysis.
Correcting the {\cPqb} jet identification and misidentification efficiencies in simulated samples affects the final \mT shapes,
so the related uncertainties are considered as shape uncertainties~\cite{Sirunyan:2017ezt}.

The systematic uncertainty due to the pileup modeling is obtained by shifting the mean of
the total inelastic $\Pp\Pp$ production cross section by $\pm 5\%$ around its nominal value~\cite{Aaboud:2016mmw},
and propagating the difference to the final \mT distributions as a shape uncertainty.

The uncertainty in the measurement of the integrated luminosity is $2.5\%$~\cite{CMS-PAS-LUM-17-001}.

The uncertainties related to the \jetToTauh background measurement in the hadronic final state are included. The statistical uncertainties in the data and simulated samples used to determine the fake factors are propagated into the final \mT distributions as a normalization uncertainty.
The limited statistical precision of samples in the signal and control region after all selections can lead to a difference in \mT shapes between the two regions. This effect is estimated and incorporated as a shape uncertainty.
As the \jetToTauh background is estimated by subtracting simulated events (electroweak/top contribution) from the control data sample,
all uncertainties related to the simulated samples are propagated to this background.
These uncertainties are scaled to correspond to the contribution from simulated events in the control region after all selections,
and anticorrelated between the \jetToTauh background and the other background processes.

The reference cross sections used to normalize each simulated background process are varied within
their theoretical uncertainties related to the choice of renormalization and factorization (RF) scales and PDFs~\cite{Butterworth:2015oua}.
For \ttbar and single top quark processes, the effect of \mtop on the cross sections is considered
by varying \mtop by 1.0\GeV around the nominal value of 172.5\GeV.
Theoretical uncertainties in the acceptance of signal and background events
are determined by varying the RF scales
and PDFs~\cite{Butterworth:2015oua}.
For the RF uncertainties, the RF scales are varied by factors of 0.5 and 2, excluding the extreme variations where one scale is varied by 0.5 and the other one by 2. The envelope of the six variations is used to determine the total uncertainty.
The cross section and acceptance uncertainties are uncorrelated between different background processes.

The uncertainty arising from the parton shower modeling is included for the dominant \ttbar background in the leptonic final states.
Four parton showering variations are included by perturbing the initial- and final-state parameters~\cite{Skands:2014pea}, the matching of jets from matrix element calculations and from parton shower, and the underlying event tune~\cite{CMS:2016kle}. The parton shower uncertainties are derived in each category and are applied as normalization uncertainties, uncorrelated between categories.
The leptonic final states are sensitive to the parton shower modeling due to the event categorization based on the jet multiplicity.
In the hadronic final state, the event selection is inclusive in jet multiplicity and thus this uncertainty is neglected.

For the intermediate-mass signal samples, an additional normalization uncertainty is assigned to incorporate the statistical uncertainties of the samples used in the calculation of the LO-to-NLO correction factors.

The statistical uncertainties related to the finite number of events in the final \mT distributions are taken into account using the Barlow--Beeston method~\cite{Barlow:1993dm}.

\section{Results}
\label{sec:results}

A simultaneous binned maximum likelihood fit is performed over all the categories in the three final states.
In total, 36 \mT distributions (two from the \tauhjets final state, 18 from the \ltauh final state, and 16 from the \lnotauh final state) are fitted.
The distributions are binned according to the statistical precision of the samples, separately for each category.
This leads to wider bins in the tail of the distributions, such that the last bin extends to 5\TeV.
The systematic uncertainties are incorporated as nuisance parameters in the likelihood.
They are profiled in the fit according to their probability density functions, taking correlations into account.
For normalization uncertainties, log-normal probability density functions are used as priors.
For shape uncertainties, polynomial interpolation is used to derive continuous prior distributions from the nominal and varied \mT shape templates.
The expected event yields after a background-only fit to the data and the observed yields are summarized in Table~\ref{tab:yields}.

\begin{table}
	\centering
	\topcaption{Number of expected and observed events for the three final states after all selections, summed over all categories in each final state.
	For background processes, the event yields after a background-only fit and the corresponding uncertainties are shown.
	For the \Hpm mass hypotheses of 100, 200, and 2000\GeV, the signal yields are normalized to an \Hpm production cross section of 1\pb and the total systematic uncertainties (prior to the fit) are shown.}
	\label{tab:yields}

\renewcommand{\arraystretch}{1.1}
  \begin{tabular}{ l l l l }
  Process & \tauhjets & \ltauh & \lnotauh \\
  \hline
  Jets misid. as \tauh       & $4619 \pm 35$  & \NA & \NA \\
  \ttbar                                    & $1455 \pm 13$  & $30560 \pm 470   $ & $174740 \pm 350$ \\
  Single \cPqt                            & $801  \pm 13$  & $3006  \pm 49    $ & $26130  \pm 260$ \\
  Electroweak                        & $1739 \pm 18$  & $2760  \pm 37    $ & $52310  \pm 220$ \\
  [\cmsTabSkip]
  Total expected from the SM                & $8614 \pm 42$  & $36320 \pm 500   $ & $253190 \pm 400$ \\
  Observed                                 & $8647           $  & $36277               $ & $253236            $ \\
  [\cmsTabSkip]
  \Hpm signal, $m_{\Hpm} = 100\GeV$          & $20  \pm 3           $  & $160 \pm 20 $      &         $ 241 \pm 26       $ \\
  \Hpm signal, $m_{\Hpm} = 200\GeV$          & $156 \pm 22       $  & $327 \pm 37$      &         $ 682 \pm 61       $ \\
  \Hpm signal, $m_{\Hpm} = 2000\GeV$          & $1630  \pm 310    $  & $369 \pm 24$      &         $ 1571 \pm 99      $ \\
  \hline
  \end{tabular}
\renewcommand{\arraystretch}{1.0}
\end{table}

\begin{figure}[hbp]
	\centering
	\includegraphics[width=0.49\textwidth]{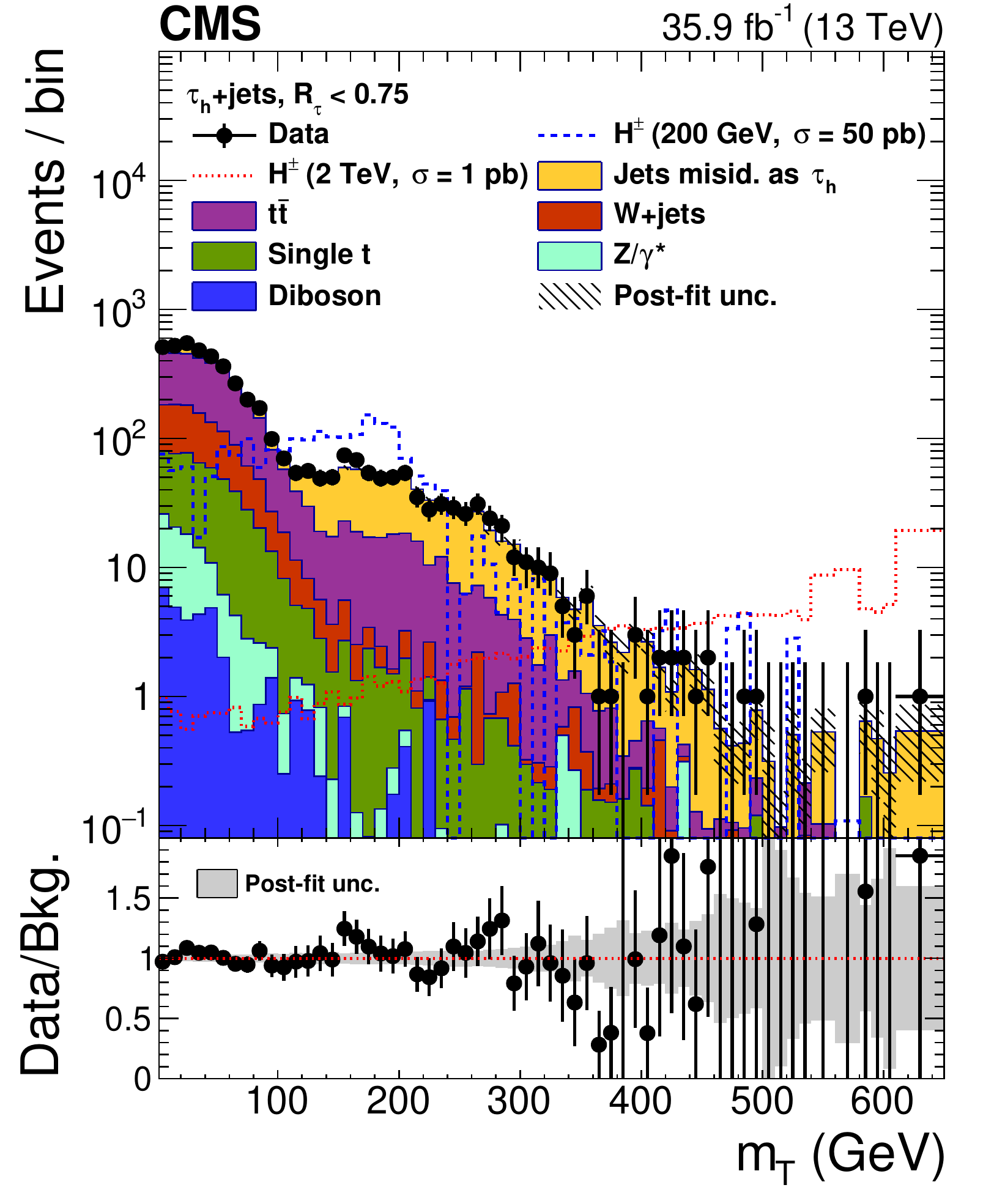}
	\includegraphics[width=0.49\textwidth]{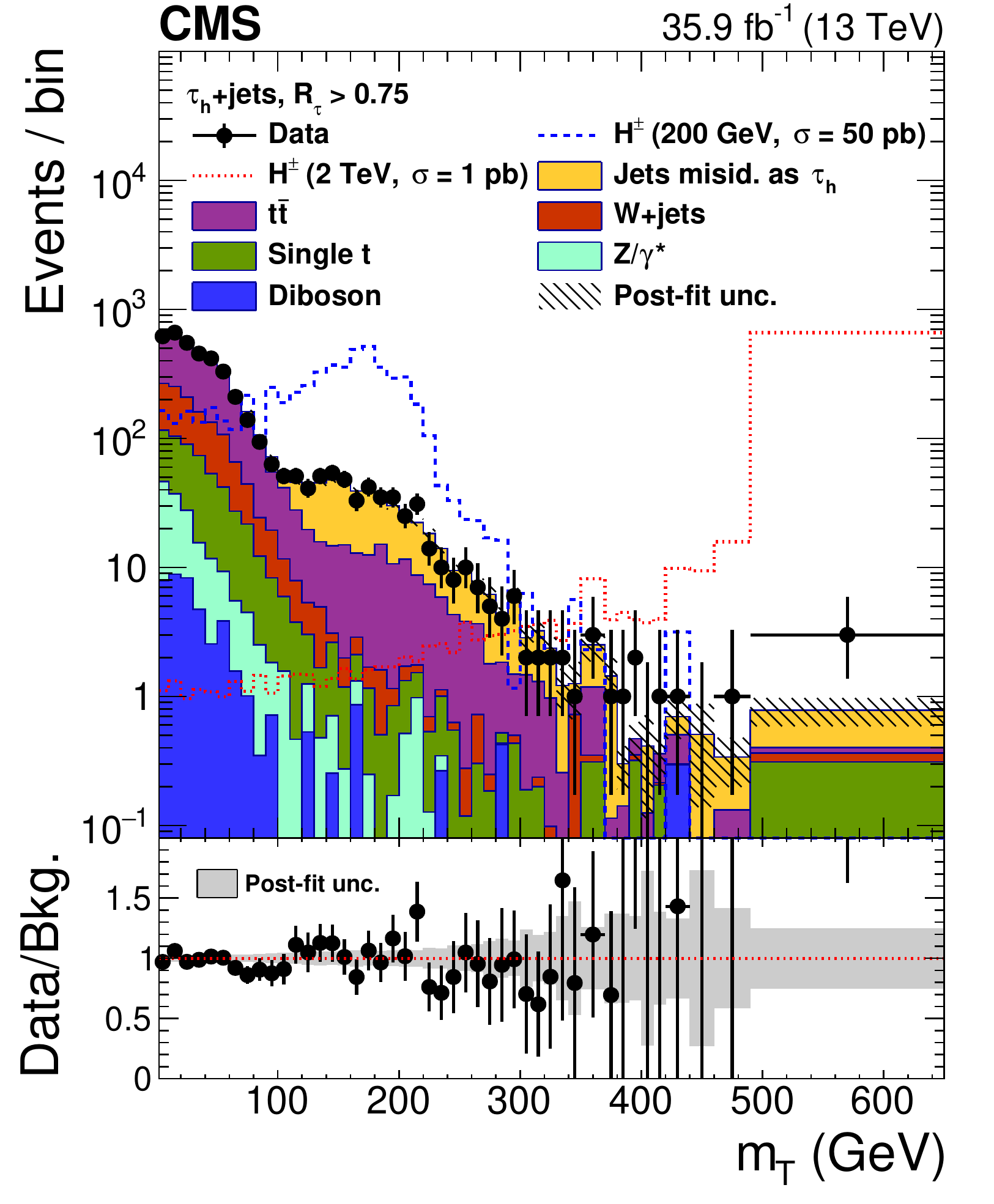}
	\caption{The transverse mass distributions in the \tauhjets final state after a background-only fit to the data. Left: category defined by \RtauLess. Transverse mass values up to $5\TeV$ are considered in the fit, but the last bins with $\mT > 650\GeV$ do not contain any observed events.
Right: category defined by \RtauMore. The last bin shown extends to $5\TeV$.}
	\label{fig:res:mt_had}
\end{figure}

\begin{figure}[hbp]
	\centering
	\includegraphics[width=0.49\textwidth]{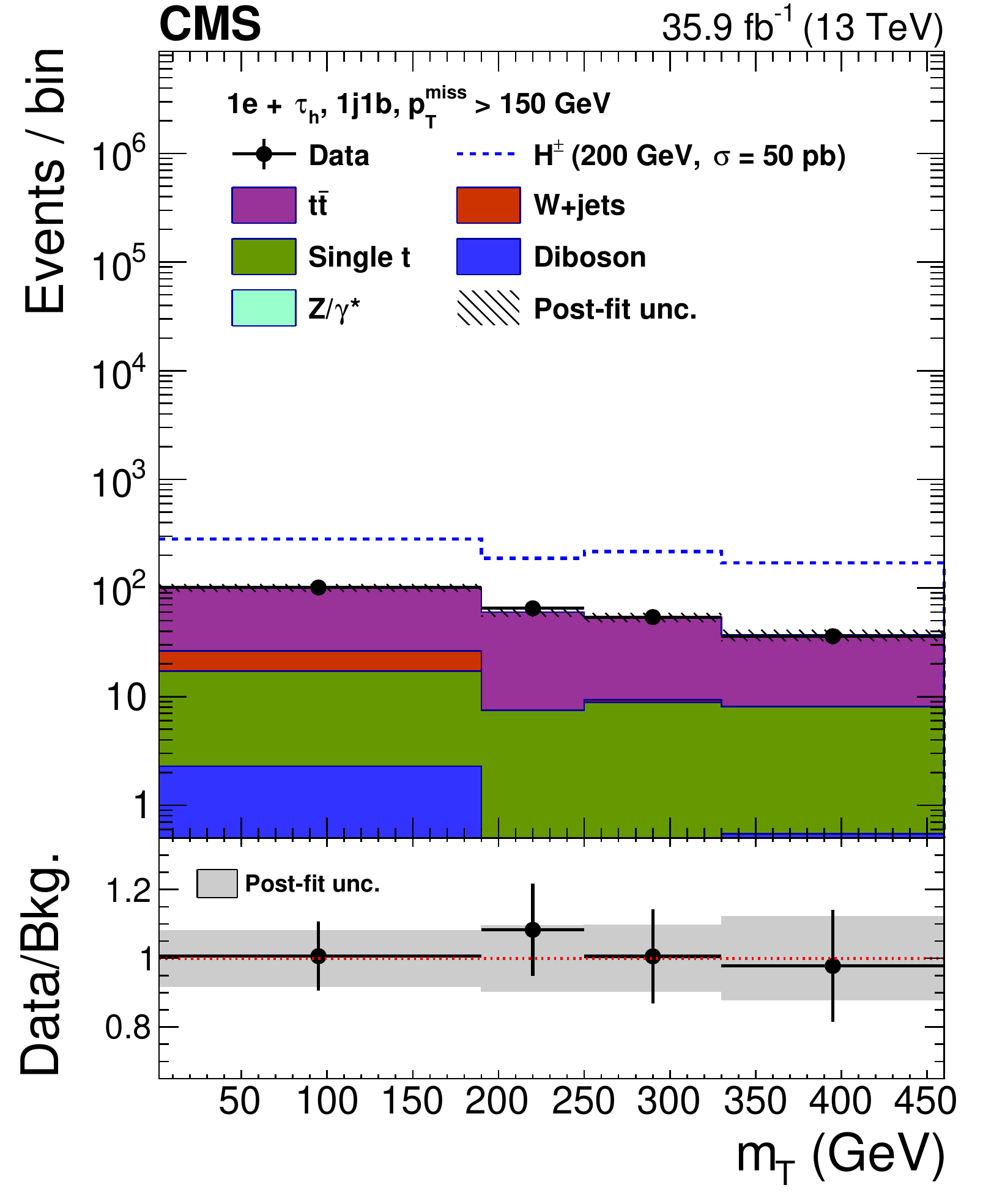}
	\includegraphics[width=0.49\textwidth]{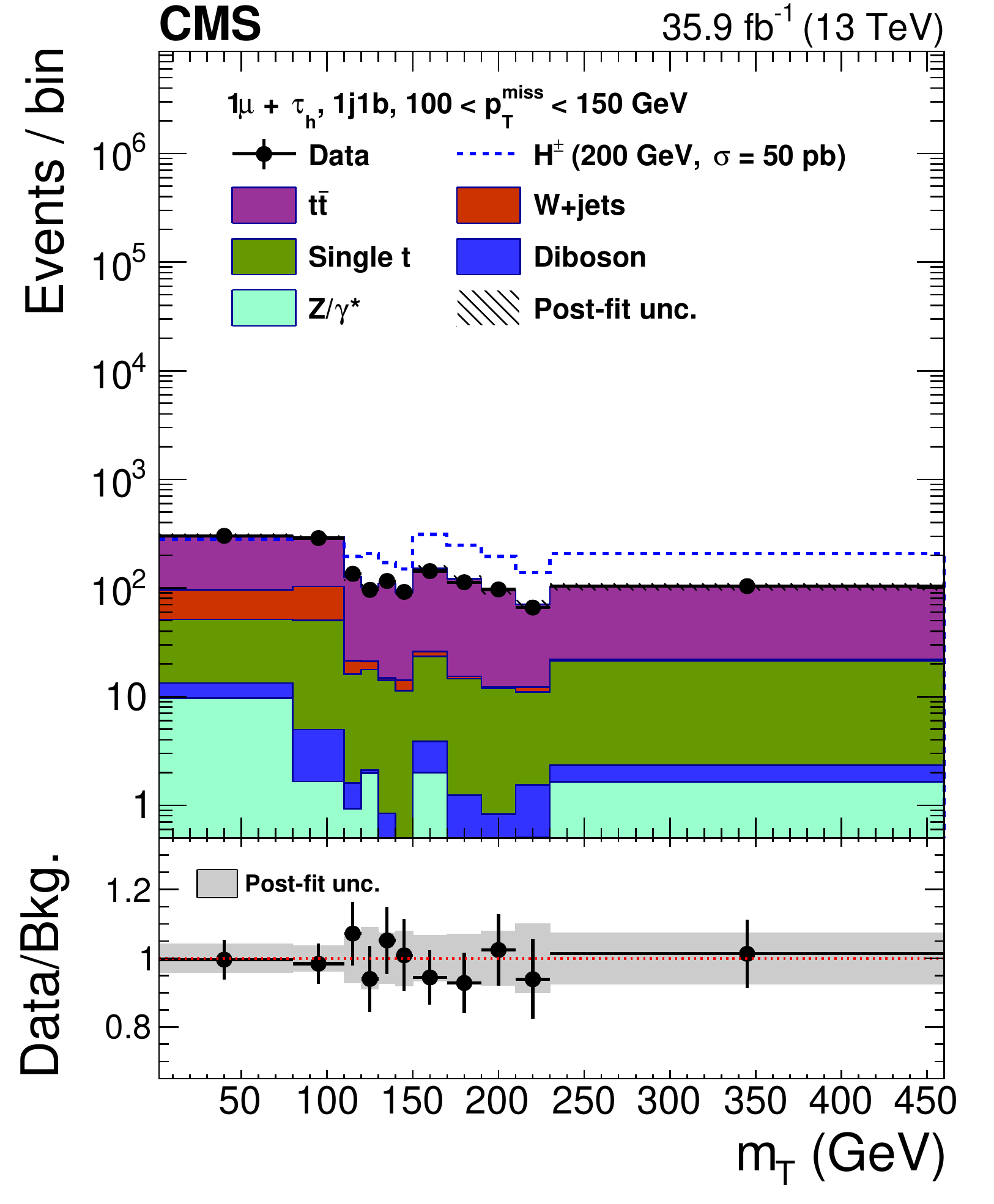}
	\caption{The transverse mass distributions for two \ltauh categories with high signal sensitivity after a background-only fit to the data. Left: category with one electron, one \tauh, one jet identified as a {\cPqb} jet, and $\ptmiss>150\GeV$. Right: category with one muon, one \tauh, one jet identified as a {\cPqb} jet and $100<\ptmiss<150\GeV$. In both categories, the last bin shown extends to $5\TeV$.}
	\label{fig:res:mt_lep}
\end{figure}

\begin{figure}[hbp]
	\centering
	\includegraphics[width=0.49\textwidth]{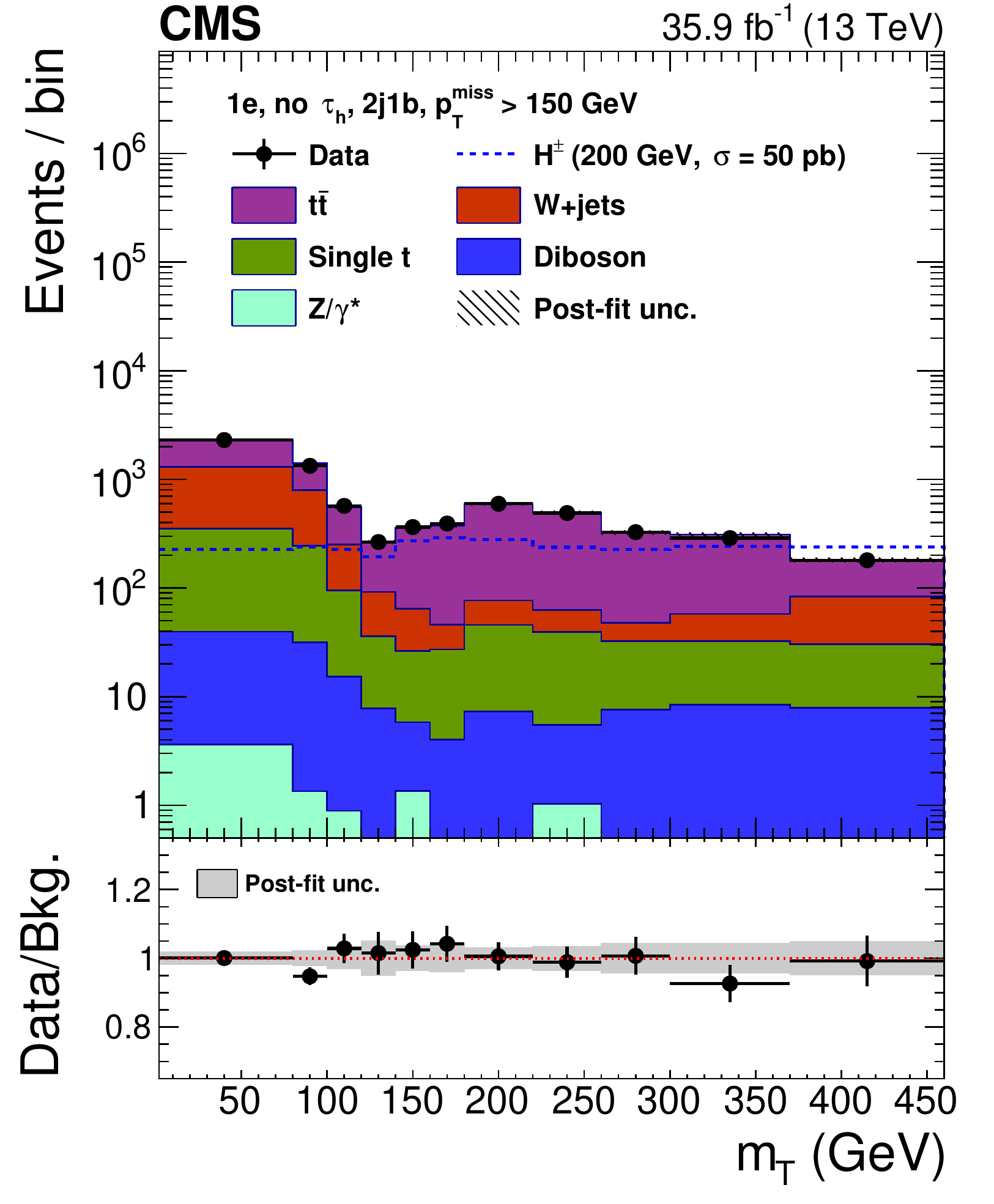}
	\includegraphics[width=0.49\textwidth]{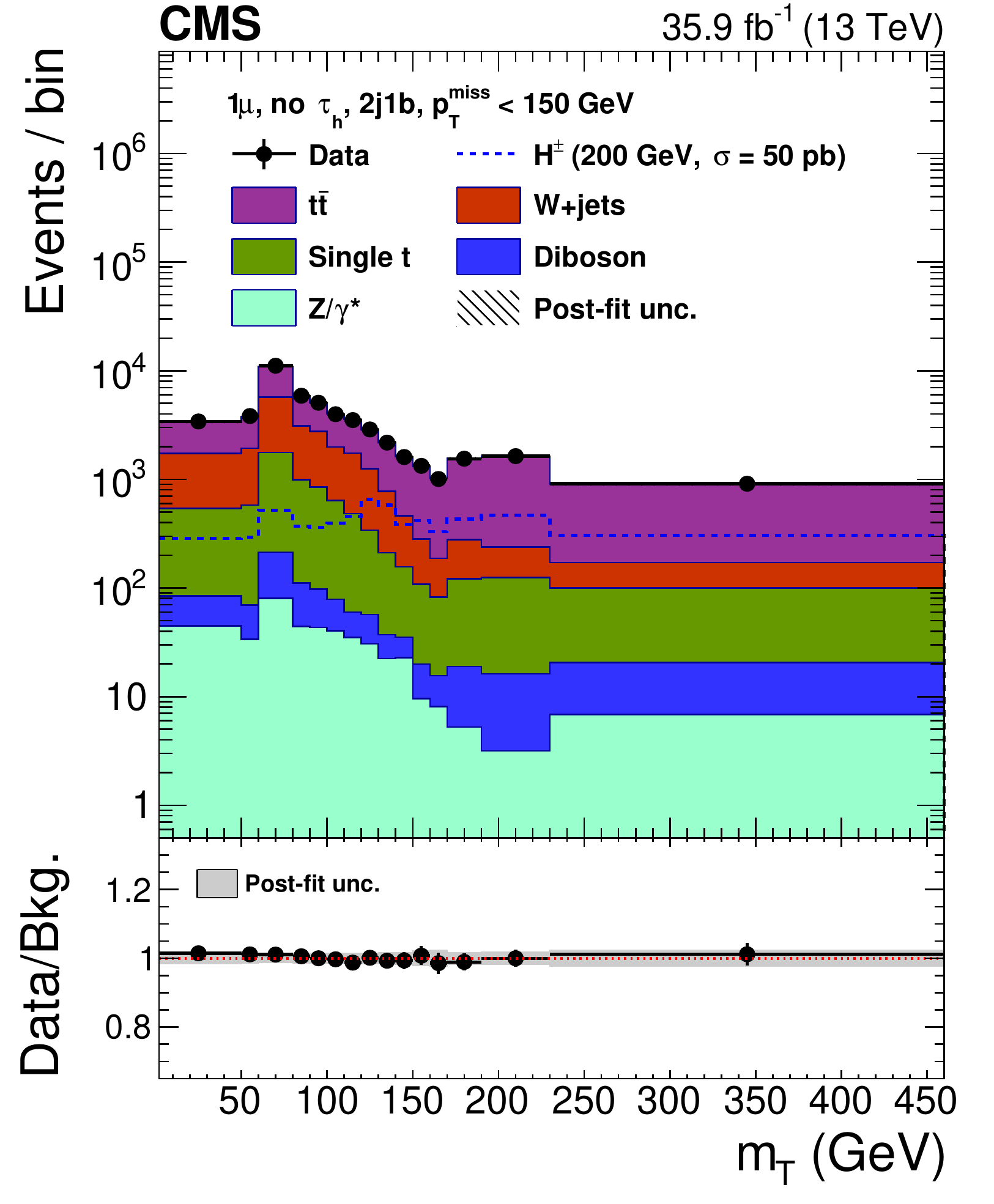}
	\caption{The \mT distributions for two \lnotauh categories with high signal sensitivity after a background-only fit to the data. Left: category with one electron, no \tauh, two jets (one identified as a {\cPqb} jet), and $\ptmiss>150\GeV$. Right: category with one muon, no \tauh, two jets (one identified as a {\cPqb} jet) and $\ptmiss<150\GeV$. In both categories, the last bin shown extends to $5\TeV$.}
	\label{fig:res:mt_lep_notauh}
\end{figure}

The distributions of \mT after a background-only fit to the data
are shown in Fig.~\ref{fig:res:mt_had} for both categories in the \tauhjets final state, in Fig.~\ref{fig:res:mt_lep} for 
two categories with high signal sensitivity in the \ltauh final state,
and in Fig.~\ref{fig:res:mt_lep_notauh} for two high-sensitivity categories
in the \lnotauh final state.
No significant excess is observed in any of the categories, and the result of the simultaneous fit is found to agree with the SM prediction.

The modified frequentist \CLs criterion~\cite{CLS1,CLS2} based on the profile likelihood ratio test statistic~\cite{ATL-PHYS-PUB-2011-011} is applied to determine the 95\% confidence level (\CL) limit for the product of the \Hpm production cross section and the branching fraction \BHpmtaunu.
The asymptotic approximation~\cite{Cowan:2010js} is used throughout the analysis. 
Pseudo-experiments are performed for selected signal mass hypotheses to verify the validity of the asymptotic approximation. 
For the \Hpm mass range up to 165\GeV, the limit on $\BtopToHpm\BHpmtaunu$ is calculated, scaling down the \ttbar background component consistently with the \BtopToHpm signal hypothesis, and the result is interpreted as a limit on \heavyLimit  by assuming
$\sigmaHplus = 2\sigma_{\ttbar}\BtopToHpm(1-\BtopToHpm)$,
where the \ttbar production cross section $\sigma_{\ttbar}$ is assumed unmodified by the presence of \Hpm and the value of
$831.76\unit{pb}$ is used~\cite{Czakon:2011xx,Kidonakis:2013zqa}.
For the \Hpm mass range from 170\GeV to 3\TeV, the limit on \heavyLimit is calculated without assuming a specific production mode.

\begin{figure}[h]
	\centering
	\includegraphics[width=0.49\textwidth]{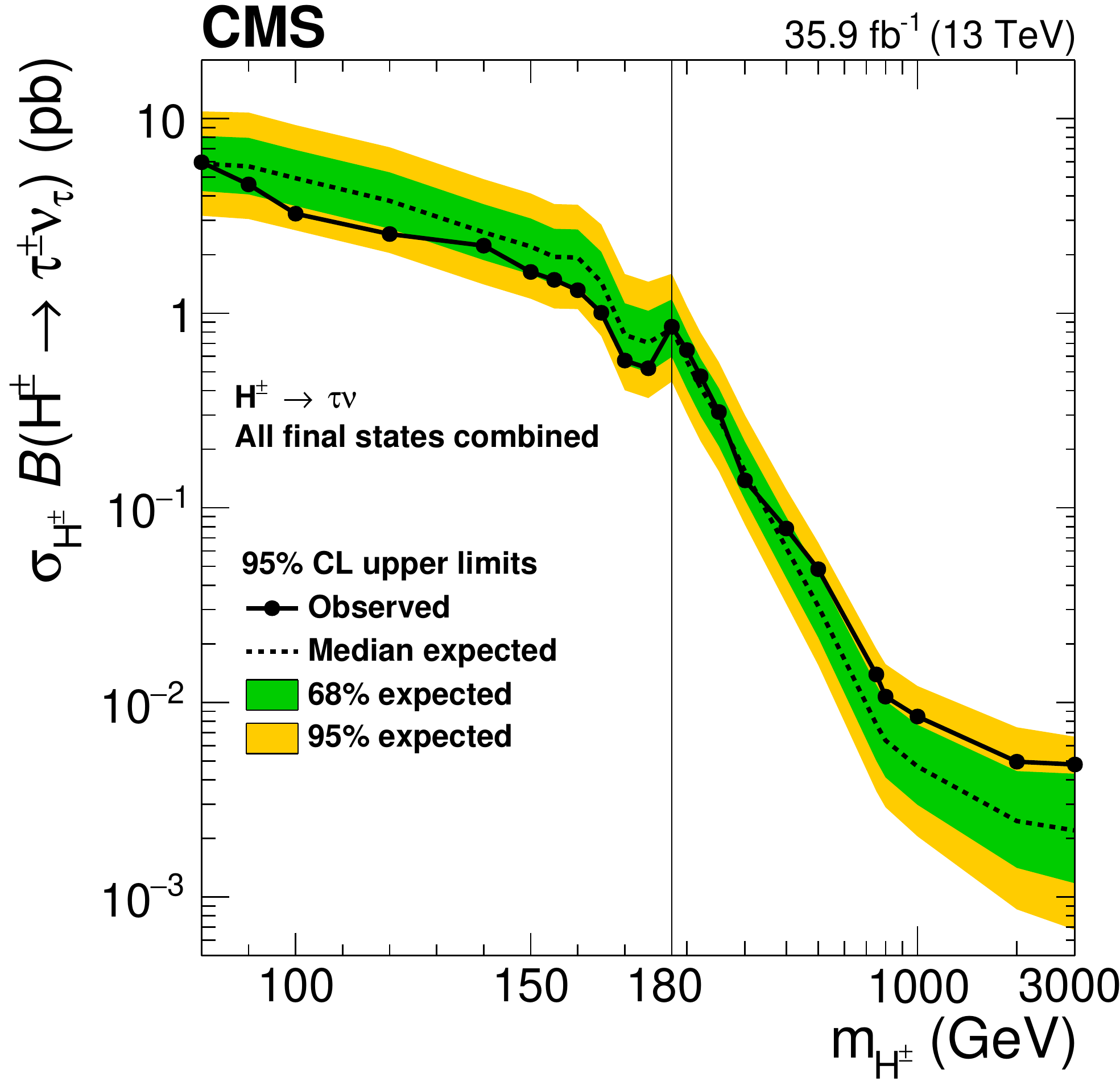}
	\includegraphics[width=0.49\textwidth]{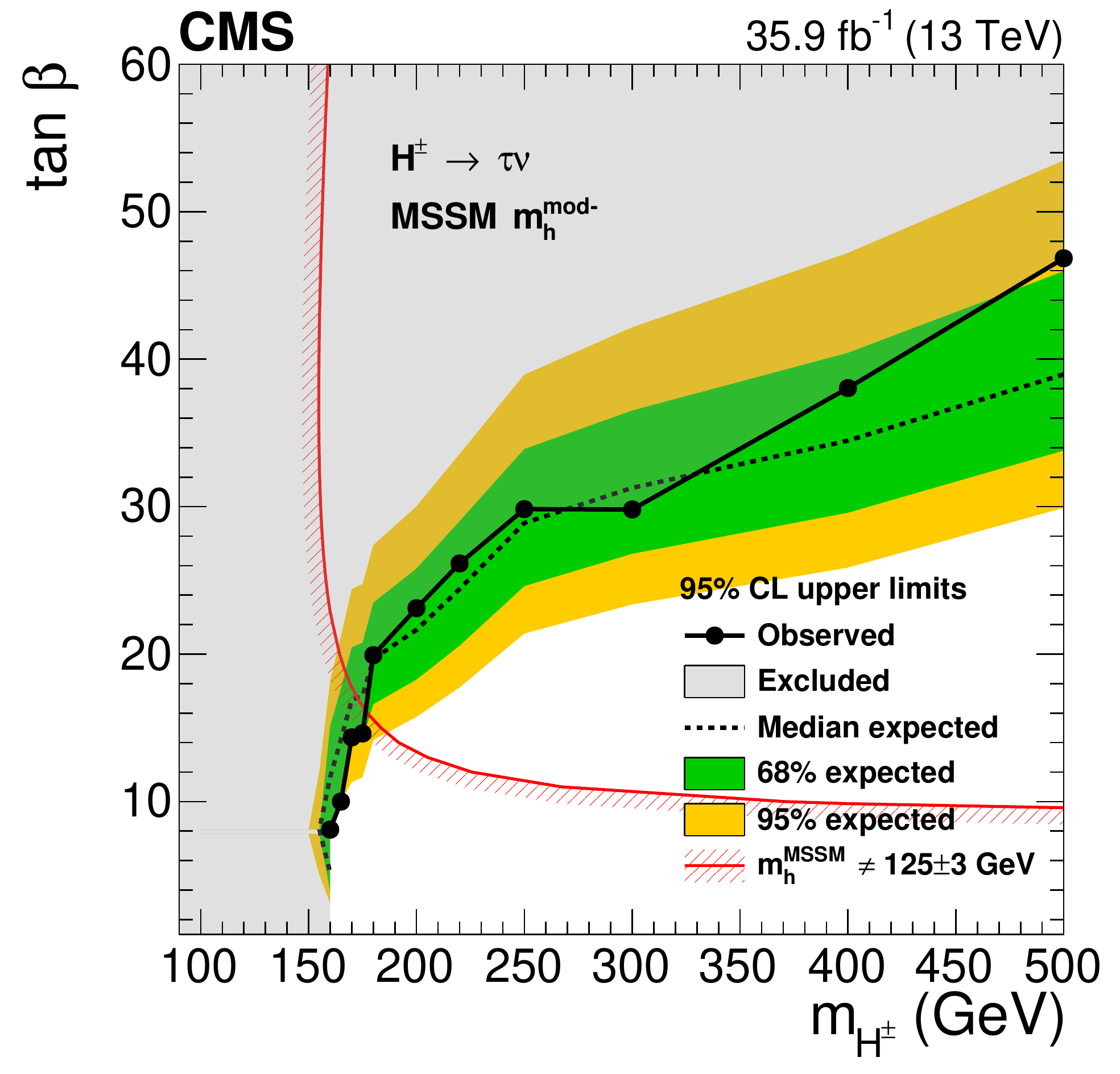}
	\caption{The observed 95\% \CL exclusion limits on \heavyLimit (solid black points),
	compared to the expected limit assuming only standard model processes (dashed line) for the \Hpm mass range from 80\GeV to 3\TeV (left),
	and the same limit interpreted in the \mhmodm benchmark scenario (right).
	The green (yellow) bands represent one (two) standard deviations from the expected limit.
	On the left, the horizontal axis is linear from 80 to 180\GeV and logarithmic for larger \mHpm values.
    On the right, the region below the red line is excluded assuming that the observed neutral Higgs boson
	is the light CP-even 2HDM Higgs boson with a mass of ${125\pm 3\GeV}$,
	where the uncertainty is the theoretical uncertainty in the mass calculation.}
	\label{fig:res:combined}
\end{figure}

The model-independent upper limit with all final states and categories combined is shown on the left side of Fig.~\ref{fig:res:combined}.
The numerical values are listed in Table~\ref{tab:limits}. The observed limit ranges from 6\pb at 80\GeV to 5\fb at 3\TeV.
For the light \Hpm mass range of 80--160\GeV, the limit corresponds to
$\BtopToHpm\BHpmtaunu$ values between 0.36\% (at 80\GeV) and 0.08\% (at 160\GeV).
In the light \Hpm mass range, this is the most stringent limit on $\BtopToHpm\BHpmtaunu$ to date set by the CMS Collaboration,
with a factor of 1.5--3.0 improvement with respect to Ref.~\cite{Khachatryan:2015qxa}, depending on \mHpm.
In the intermediate mass range of 165--175\GeV, this is the first limit on \heavyLimit set by the CMS Collaboration.
The drop in the expected and observed limits in the intermediate region
is not predicted from theory~\cite{Degrande:2016hyf} but is rather an experimental feature
explained by the fact that in this region LO signal samples are used instead of NLO.
This dip is mitigated but not completely cancelled by the LO-to-NLO corrections extrapolated from the surrounding mass regions.
In the heavy mass range from 180\GeV, this result extends the search region up to $\mHpm = 3\TeV$, compared to 600\GeV in Ref.~\cite{Khachatryan:2015qxa}.

In the light and intermediate \Hpm mass regions all three final states contribute significantly to the sensitivity, 
and the combined limits are on average {$\approx$}40\% lower compared to the \tauhjets final state alone. 
In the heavy \Hpm mass region, the sensitivity of the leptonic final states 
decreases, and the \tauhjets final state starts to dominate the limit as \mHpm increases. 
Above $\mHpm = 500\GeV$ the combined limit is solely driven by the \tauhjets final state. 

The limit is interpreted in the MSSM \mhmodm benchmark scenario~\cite{Carena:2013ytb}
by comparing the observed limit on the \Hpm cross section to the theoretical cross sections predicted in this scenario~\cite{deFlorian:2016spz,Flechl:2014wfa,Degrande:2015vpa,Dittmaier:2009np,Berger:2003sm,Degrande:2016hyf}.
The MSSM \mhmodm scenario is specified using low-energy MSSM parameters and is designed to give a mass of approximately 125\GeV
for the light CP-even Higgs boson over a wide region of the parameter space.
The limit for the MSSM \mhmodm scenario in the $\mHpm$--$\tanbeta$ plane is shown on the right side of Fig.~\ref{fig:res:combined}.
Based on the observed limit, all values of the parameter \tanbeta from 1 to 60 are excluded for \mHpm values up to 160\GeV. 
The limit extends to $\mHpm = 500\GeV$. 
For $\mHpm = 200$ (400)\GeV, the observed limit excludes all \tanbeta values above 26 (40), compared to 45 (56) excluded in Ref.~\cite{Khachatryan:2015qxa}. 

\begin{table}
	\centering
	\topcaption{The expected and observed 95\% \CL exclusion limits on \heavyLimit
	for the \Hpm mass range from 80\GeV to 3\TeV.
	The $\pm 1$ s.d. ($\pm 2$ s.d.) refers to one (two) standard deviations from the expected limit.}
	\label{tab:limits}
\renewcommand{\arraystretch}{1.1}
\begin{tabular}{ c c c c c c c }
\mHpm & \multicolumn{5}{ c }{Expected limit (pb)} & Observed \\ \cline{2-6}
(\GeVns{})   & $-2$ s.d.  & $-1$ s.d. & median & +1 s.d. & +2 s.d. & limit (\unit{pb}) \\
\hline
80 & 3.17 & 4.25 & 5.87 & 8.15 & 10.89 & 5.97 \\
90 & 3.05 & 4.08 & 5.69 & 7.96 & 10.75 & 4.59 \\
100 & 2.67 & 3.56 & 4.94 & 6.90 & 9.26 & 3.24 \\
120 & 2.04 & 2.72 & 3.78 & 5.29 & 7.12 & 2.55 \\
140 & 1.41 & 1.87 & 2.61 & 3.63 & 4.88 & 2.22 \\
150 & 1.19 & 1.58 & 2.20 & 3.07 & 4.14 & 1.63 \\
155 & 1.06 & 1.41 & 1.95 & 2.71 & 3.64 & 1.48 \\
160 & 1.05 & 1.39 & 1.93 & 2.69 & 3.61 & 1.31 \\
165 & 0.76 & 1.02 & 1.45 & 2.67 & 2.86 & 1.01 \\
170 & 0.40 & 0.54 & 0.77 & 1.12 & 1.59 & 0.57 \\
175 & 0.37 & 0.50 & 0.71 & 1.03 & 1.45 & 0.52 \\
180 & 0.44 & 0.60 & 0.83 & 1.18 & 1.59 & 0.85 \\
200 & 0.30 & 0.41 & 0.57 & 0.80 & 1.09 & 0.65 \\
220 & 0.22 & 0.30 & 0.41 & 0.58 & 0.80 & 0.47 \\
250 & 0.15 & 0.21 & 0.29 & 0.41 & 0.56 & 0.31 \\
300 & 0.08 & 0.11 & 0.15 & 0.22 & 0.30 & 0.14 \\
400 & 0.032 & 0.043 & 0.062 & 0.090 & 0.125 & 0.078 \\
500 & 0.016 & 0.022 & 0.031 & 0.046 & 0.067 & 0.048 \\
750 & 0.0035 & 0.0050 & 0.0077 & 0.012 & 0.019 & 0.014 \\
800 & 0.0029 & 0.0041 & 0.0064 & 0.0102 & 0.0157 & 0.0107 \\
1000 & 0.0020 & 0.0030 & 0.0047 & 0.0077 & 0.0121 & 0.0085 \\
2000 & 0.0009 & 0.0014 & 0.0025 & 0.0044 & 0.0074 & 0.0050 \\
2500 & 0.0007 & 0.0012 & 0.0022 & 0.0042 & 0.0068 & 0.0047 \\
3000 & 0.0007 & 0.0012 & 0.0022 & 0.0043 & 0.0067 & 0.0048 \\
\hline
\end{tabular}
\renewcommand{\arraystretch}{1.0}
\end{table}

\section{Summary}
\label{sec:conclusion}

A search is presented for charged Higgs bosons decaying as \Hpmtaunu, using events
recorded by the CMS experiment in 2016 at a center-of-mass energy of 13\TeV.
Transverse mass distributions are reconstructed in hadronic and leptonic final states and are found to agree with the standard model expectation.
Upper limits for the product of the \Hpm production cross section and the branching fraction to \taunu are set at 95\% confidence level for an \Hpm mass ranging from 80\GeV to 3\TeV,
including the range close to the top quark mass.
The observed limit ranges from $6\pb$ at 80\GeV to $5\fb$ at 3\TeV.
The results are interpreted as constraints in the parameter space of the minimal supersymmetric standard model \mhmodm benchmark scenario.
In this scenario, all \tanbeta values from 1 to 60 are excluded for charged Higgs boson masses up to 160\GeV.

\begin{acknowledgments}

We congratulate our colleagues in the CERN accelerator departments for the excellent performance of the LHC and thank the technical and administrative staffs at CERN and at other CMS institutes for their contributions to the success of the CMS effort. In addition, we gratefully acknowledge the computing centers and personnel of the Worldwide LHC Computing Grid for delivering so effectively the computing infrastructure essential to our analyses. Finally, we acknowledge the enduring support for the construction and operation of the LHC and the CMS detector provided by the following funding agencies: BMBWF and FWF (Austria); FNRS and FWO (Belgium); CNPq, CAPES, FAPERJ, FAPERGS, and FAPESP (Brazil); MES (Bulgaria); CERN; CAS, MoST, and NSFC (China); COLCIENCIAS (Colombia); MSES and CSF (Croatia); RPF (Cyprus); SENESCYT (Ecuador); MoER, ERC IUT, and ERDF (Estonia); Academy of Finland, MEC, and HIP (Finland); CEA and CNRS/IN2P3 (France); BMBF, DFG, and HGF (Germany); GSRT (Greece); NKFIA (Hungary); DAE and DST (India); IPM (Iran); SFI (Ireland); INFN (Italy); MSIP and NRF (Republic of Korea); MES (Latvia); LAS (Lithuania); MOE and UM (Malaysia); BUAP, CINVESTAV, CONACYT, LNS, SEP, and UASLP-FAI (Mexico); MOS (Montenegro); MBIE (New Zealand); PAEC (Pakistan); MSHE and NSC (Poland); FCT (Portugal); JINR (Dubna); MON, RosAtom, RAS, RFBR, and NRC KI (Russia); MESTD (Serbia); SEIDI, CPAN, PCTI, and FEDER (Spain); MOSTR (Sri Lanka); Swiss Funding Agencies (Switzerland); MST (Taipei); ThEPCenter, IPST, STAR, and NSTDA (Thailand); TUBITAK and TAEK (Turkey); NASU and SFFR (Ukraine); STFC (United Kingdom); DOE and NSF (USA).

\hyphenation{Rachada-pisek} Individuals have received support from the Marie-Curie program and the European Research Council and Horizon 2020 Grant, contract No. 675440 (European Union); the Leventis Foundation; the A. P. Sloan Foundation; the Alexander von Humboldt Foundation; the Belgian Federal Science Policy Office; the Fonds pour la Formation \`a la Recherche dans l'Industrie et dans l'Agriculture (FRIA-Belgium); the Agentschap voor Innovatie door Wetenschap en Technologie (IWT-Belgium); the F.R.S.-FNRS and FWO (Belgium) under the ``Excellence of Science - EOS" - be.h project n. 30820817; the Ministry of Education, Youth and Sports (MEYS) of the Czech Republic; the Lend\"ulet (``Momentum") Program and the J\'anos Bolyai Research Scholarship of the Hungarian Academy of Sciences, the New National Excellence Program \'UNKP, the NKFIA research grants 123842, 123959, 124845, 124850 and 125105 (Hungary); the Council of Science and Industrial Research, India; the HOMING PLUS program of the Foundation for Polish Science, cofinanced from European Union, Regional Development Fund, the Mobility Plus program of the Ministry of Science and Higher Education, the National Science Center (Poland), contracts Harmonia 2014/14/M/ST2/00428, Opus 2014/13/B/ST2/02543, 2014/15/B/ST2/03998, and 2015/19/B/ST2/02861, Sonata-bis 2012/07/E/ST2/01406; the National Priorities Research Program by Qatar National Research Fund; the Programa Estatal de Fomento de la Investigaci{\'o}n Cient{\'i}fica y T{\'e}cnica de Excelencia Mar\'{\i}a de Maeztu, grant MDM-2015-0509 and the Programa Severo Ochoa del Principado de Asturias; the Thalis and Aristeia programs cofinanced by EU-ESF and the Greek NSRF; the Rachadapisek Sompot Fund for Postdoctoral Fellowship, Chulalongkorn University and the Chulalongkorn Academic into Its 2nd Century Project Advancement Project (Thailand); the Welch Foundation, contract C-1845; and the Weston Havens Foundation (USA).
\end{acknowledgments}
\bibliography{auto_generated}
\cleardoublepage \appendix\section{The CMS Collaboration \label{app:collab}}\begin{sloppypar}\hyphenpenalty=5000\widowpenalty=500\clubpenalty=5000\vskip\cmsinstskip
\textbf{Yerevan Physics Institute, Yerevan, Armenia}\\*[0pt]
A.M.~Sirunyan, A.~Tumasyan
\vskip\cmsinstskip
\textbf{Institut f\"{u}r Hochenergiephysik, Wien, Austria}\\*[0pt]
W.~Adam, F.~Ambrogi, E.~Asilar, T.~Bergauer, J.~Brandstetter, M.~Dragicevic, J.~Er\"{o}, A.~Escalante~Del~Valle, M.~Flechl, R.~Fr\"{u}hwirth\cmsAuthorMark{1}, V.M.~Ghete, J.~Hrubec, M.~Jeitler\cmsAuthorMark{1}, N.~Krammer, I.~Kr\"{a}tschmer, D.~Liko, T.~Madlener, I.~Mikulec, N.~Rad, H.~Rohringer, J.~Schieck\cmsAuthorMark{1}, R.~Sch\"{o}fbeck, M.~Spanring, D.~Spitzbart, W.~Waltenberger, J.~Wittmann, C.-E.~Wulz\cmsAuthorMark{1}, M.~Zarucki
% ...
\vskip\cmsinstskip
% ...
1:  Also at Vienna University of Technology, Vienna, Austria\\
% ...
\end{sloppypar}
\end{document}